\documentstyle[12pt]{article}

\catcode`\@=11
\@addtoreset{equation}{section}

\def\l#1{{\lambda}_{#1}}
\def\lp#1{{\lambda^{\prime}_{#1}}}
\def\lpp#1{{\lambda^{\prime\prime}_{#1}}}
\def\C#1{{C}_{#1}}
\def\Cp#1{{C^{\prime}_{#1}}}
\def\Cpp#1{{C^{\prime\prime}_{#1}}}

\begin{document}

\baselineskip 24pt

\begin{titlepage}
\begin{flushright}
SUSX-TH/96-003, OUTP-96-01P, IEM-FT-125/96\\
hep-ph/9602381
\end{flushright}
\vspace{.2in}
\begin{center}

{\large{\bf R-parity Violation Effects through
Soft Supersymmetry Breaking Terms
and the Renormalisation Group}}
\bigskip \\
B.~de Carlos$^1$ and P.~L.~White$^2$
\\ \mbox{} \\
{\it 
$^1$School of Mathematical and Physical Sciences, 
University of Sussex, \\ Falmer, Brighton BN1 9QH, UK. \\
Email: {\tt B.De-Carlos@sussex.ac.uk} \\
and \\
$^2$Theoretical Physics, University of Oxford, \\
1 Keble Road, Oxford OX1 3NP, UK. \\
Email: {\tt plw@thphys.ox.ac.uk}
}
\\ \vspace{.5in}
{\bf Abstract} \smallskip \end{center} \setcounter{page}{0}
We present full renormalisation group equations for the MSSM with
R-parity violation, including all soft supersymmetry breaking terms.
The inclusion of dimensionless R-parity violating couplings can
generate many possible low energy effects through the generation of
off-diagonal soft masses violating lepton and quark flavour, and
through the generation of lepton-Higgs mixing. We discuss the
relation between the weak and unification scale R-parity violation
and study the effects on neutrino mass generation and $\mu\to e\gamma$.

\end{titlepage}

\section{Introduction}
Among the different proposals to extend the Standard Model (SM),
supersymmetric models are probably the most attractive ones as their
fundamental ingredient, supersymmetry (SUSY) \cite{revs}, provides a
very  elegant solution to the so--called hierarchy problem. However the
most popular SUSY extension of the SM is not the most general one, as it
is commonly assumed that a discrete symmetry, known as R--parity,
forbids all the baryon ($B$) and lepton ($L$) number violating
couplings, hence avoiding possible rapid proton decay. To each field is
assigned an R-parity given by $R\equiv (-1)^{3B+L+2S}$ ($S$ being the
spin) so that all the SM particles have $R=+1$ while their SUSY
partners have $R=-1$. Immediate consequences of imposing such a 
symmetry are that SUSY particles are produced in pairs and that the 
lightest SUSY particle (LSP) is stable, therefore becoming a dark 
matter candidate.

However, there is no obvious reason why $B$ or $L$ violating
interactions should not be allowed separately to break R-parity
\cite{rpv,suzuki}, and the apparent constraints through the requirement
that any new physics should not ruin baryogenesis in the early universe
have been shown to be very weak \cite{cosmology}. The corresponding
terms in the superpotential are:
\begin{equation}
W_{\hbox{$\scriptstyle B\!\!\! /$}} = \frac{1}{2} \lpp{ijk}{u}_i{d}_j{d}_k 
\label{bv}
\end{equation}
and 
\begin{equation}
W_{\hbox{$\scriptstyle L\!\!\! /$}}
 = \frac{1}{2} \l{ijk}L_iL_j{e}_k + \lp{ijk}L_iQ_j{d}_k \; \;,
\label{lv}
\end{equation}
where $u$, $d$, $e$ stand for the u-type, d-type and lepton singlet 
superfields, and $Q$, $L$ are the quark and lepton doublet superfields
respectively, which generate dimensionless couplings in the
Lagrangian, given by:
\begin{equation}
{\cal L}_{\hbox{$\scriptstyle B\!\!\! /$}} = \frac{1}{2} \lpp{ijk} 
( u^{c}_{i}d^{c}_{j}\tilde{d}^{*}_{k} +
      u^c_{i}\tilde{d}^{*}_{j}d^c_{k} + 
    \tilde{u}^{*}_{i} d^c_{j} d^c_{k} ) + {\rm h.c.}
\label{lbv}
\end{equation}
and:
\begin{eqnarray}
{\cal L}_{\hbox{$\scriptstyle L\!\!\! /$}} & = & \frac{1}{2} \l{ijk} \left( 
     \bar{\nu}^c_{Li}e_{Lj}\tilde{e}^{*}_{Rk}  - 
e_{Li}\bar{\nu}^c_{Lj}\tilde{e}^{*}_{Rk}
+ \nu_{Li}\tilde{e}_{Lj} \bar{e}_{Rk} - e_{Li}\tilde{\nu}_{Lj}\bar{e}_{Rk}
      \right . \cr
     && \qquad + \left. \tilde{\nu}_{Li}e_{Lj}\bar{e}_{Rk} - 
\tilde{e}_{Li}\nu_{Lj}\bar{e}_{Rk}
           \right) + {\rm h.c.}  
     \label{llv} \\
     & + & \lp{ijk}\left( \bar{\nu}^c_{Li} d_{Lj}\tilde{d}^{*}_{Rk}
     - \bar{e}^c_{Ri}u_{Lj} \tilde{d}^{*}_{Rk} + 
\nu_{Li}\tilde{d}_{Lj}\bar{d}_{Rk}
     - e_{Li}\tilde{u}_{Lj}\bar{d}_{Rk} \right. \cr
     && \qquad + \left. \tilde{\nu}_{Li}d_{Lj}\bar{d}_{Rk}
     - \tilde{e}_{Li}u_{Lj}\bar{d}_{Rk} \right ) + {\rm h.c.} \cr \nonumber 
\end{eqnarray}
Besides these, we also expect dimensionful terms analogous to the
R--parity conserving {\em soft breaking terms}; that is, we get soft
trilinear couplings and masses which violate $B$ or $L$ :
\begin{equation}
V_{\hbox{$\scriptstyle B\!\!\! /$ soft}}
 = \frac{1}{2} \Cpp{ijk} \tilde{u}_i \tilde{d}_j \tilde{d}_k + {\rm h.c.} 
          + \sum_{a,b} m_{ab}^2\varphi_a\bar\varphi_b 
\label{bsoft}
\end{equation}
\begin{equation}
V_{\hbox{$\scriptstyle L\!\!\! /$} soft} =
     \frac{1}{2} \C{ijk} \tilde{L}_i \tilde{L}_j \tilde{e}_k
               + \Cp{ijk}\tilde{L}_i \tilde{Q}_j \tilde{d}_k + {\rm h.c.}
               + \sum_{a,b} m_{ab}^{2} \varphi_a\bar\varphi_b
               + D_i\tilde L_iH_2 +{\rm h.c.}
\label{lsoft}
\end{equation}
Here the soft masses shown include such terms as $m^2_{L_iH_1}$ which
violates $L$, and also terms violating lepton and quark flavour
symmetries (where each generation is assigned its own $L_i$ and $B_i$)
whose effects will be in addition to the contribution from the CKM
sector. For a discussion of our notation, see Appendix A.

R--parity violation has been considered by many authors in the past and
recently there has been an increasing interest in the  renormalisation
group properties of R-parity violating couplings
\cite{br,gs,herbi,roger}. This is specially interesting when we try to
embed the SM as the low energy limit of a more fundamental Grand
Unified Theory (GUT). In such a framework we have a prediction for the
values of the different couplings at the unification scale, $M_{GUT}$,
and we can therefore use their renormalisation group equations (RGEs)
to obtain the corresponding values at low energies and compare them
with the existing limits. This applies to both dimensionless and
dimensionful couplings, for which the RGEs are given (the latter for
the first time) in Appendix A.

Concerning the limits which can be imposed on these couplings, there
have been many attempts to constrain them through their impact on rare
processes in the laboratory \cite{bgh,bpw,bgnn} and it turns out that
flavour changing neutral current (FCNC) processes in particular provide
us with a very powerful tool to test physics beyond the SM. The
presence of extra particles, SUSY partners in this case, together with
lepton and quark flavour violating couplings, including here
off-diagonal soft masses, in our fundamental theory can induce
transitions forbidden in the pure SM and, moreover, strongly contrained
by experiment. Consequently an upper bound can be derived on different
products of dimensionless couplings and even on some of the
dimensionful ones. Furthermore, some of these newly generated couplings
can give rise to neutrino masses and mixings which are again
experimentally constrained.

In general we expect that once we allow the violation of lepton and
quark flavour symmetries through any operator, including the R--parity
violating dimensionless couplings shown in equations (\ref{lv}) and
(\ref{lbv}), other such operators will be generated through the RGEs.
In particular, we here study for the first time how the R--parity
violating couplings generate soft terms which can lead to large FCNC
effects in the lepton sector process $\mu\to e\gamma$.

Therefore, the aim of this paper is to present a detailed study of 
the limits which can be obtained on these R-parity violating 
couplings {\em at the GUT scale} by imposing FCNC and neutrino mass 
bounds {\em at low energy}. In particular, we have focussed on $L$ 
violating interactions, but we plan to study $B$ violation in a 
forthcoming paper.

In section~2 we discuss the dimensionless R--parity violating
couplings. We sketch analytical solutions for the case when the gauge
contribution dominate in their RGEs and obtain triviality bounds for
different values of the strong coupling $\alpha_3$. In section~3 we
clarify which basis is the most appropriate to deal with lepton and
quark flavour violation and also the lepton--Higgs (LH) mixing which
arises from R--parity violation. Section~4 is devoted to a study of
sneutrino VEVs, their generation from the dimensionful RGEs, and the
limits which can be set on them by using the present constraints on
neutrino masses. In section~5 we put limits on various R--parity
violating couplings at $M_{GUT}$ by running them to the electroweak
scale and requiring that the full SUSY contribution to the FCNC process
$\mu\rightarrow e\gamma$ be less  than its actual experimental bound.
We perform a full numerical study and give some analytical
approximations. Finally in section~6 we present our conclusions.

\section{Dimensionless Couplings}
Before turning to the dimensionful couplings, it will be helpful to
discuss what conclusions we can draw from the RGEs in Appendix A with
respect to the dimensionless ones. We start by remarking that there
have recently been a number of analyses \cite{br,gs,herbi,roger}
studying the RGEs for some or all of the dimensionless couplings. Our
RGEs are consistent with both references \cite{br} and \cite{gs}, apart
from a couple of typographical errors in the gauge contributions to the
RGEs in reference \cite{br}. More complete lists of RGEs are given in
reference \cite{herbi}, but here we disagree on a number of
coefficients, even at one loop. Our equations give a different sign for
the third term in equation (2.13) and the last term in (2.16) of that
paper. Finally, we agree with the equations of reference \cite{roger}.

While the equations are too complex to be usefully solved analytically
in general, we might expect that often many (or all) of the
R--parity violating Yukawa couplings will be tiny relative to the gauge
couplings, and hence that only the effect of the latter will be
relevant. In this limit, it is straightforward to solve the RGEs
analytically \cite{analyticRGE} as follows.
\begin{eqnarray} 
\tilde\alpha_a(t) & = &
\frac{\tilde\alpha_a(t_0)}{1-2b_a(t-t_0)\tilde\alpha_a(t_0)} 
\nonumber \\
\lambda(t) & = &
\Bigl(1-22(t-t_0)\tilde\alpha_1(t_0) \Bigr )^{c_1/22}
\Bigl(1-2(t-t_0)\tilde\alpha_2(t_0) \Bigr )^{c_2/2} \nonumber \\
&& \times \Bigl(1+6(t-t_0)\tilde\alpha_3(t_0) \Bigr )^{-c_3/6}
\nonumber \\
\lambda^{\prime}(t) & = &
\Bigl(1-22(t-t_0)\tilde\alpha_1(t_0) \Bigr )^{c_1^{\prime}/22}
\Bigl(1-2(t-t_0)\tilde\alpha_2(t_0) \Bigr )^{c_2^{\prime}/2} \\
&& \times \Bigl(1+6(t-t_0)\tilde\alpha_3(t_0) 
\Bigr )^{-c_3^{\prime}/6} \nonumber \\
\lambda^{\prime\prime}(t) & = &
\Bigl(1-22(t-t_0)\tilde\alpha_1(t_0) \Bigr )^{c_1^{\prime\prime}/22}
\Bigl(1-2(t-t_0)\tilde\alpha_2(t_0) \Bigr )^{c_2^{\prime\prime}/2}
\nonumber \\
&& \times \Bigl(1+6(t-t_0)\tilde\alpha_3(t_0) 
\Bigr )^{-c_3^{\prime\prime}/6} \nonumber
\end{eqnarray}
Here we have used the convention that $t=\ln\mu$ where $\mu$ is the 
$\overline{MS}$ renormalisation scale, while
\begin{eqnarray}
\tilde\alpha_a & = & g_a^2/(16\pi^2) \nonumber \\
b_a & = & (11,\ 1,\ -3) \nonumber \\
c_a & = & (3,\ 3,\ 0)  \\
c_a^{\prime} & = & (7/9,\ 3,\ 16/3) \nonumber \\
c_a^{\prime\prime} & = & (4/3,\ 0,\ 8) \nonumber
\end{eqnarray}
and the label $a$ runs over gauge groups 1,2,3. Although it is possible
to derive analytical solutions even when the Yukawa couplings are large
\cite{analyticRGE}, in practice the expressions become so complicated
that they are not particularly useful.

If we solve these equations with $t=\ln(M_{GUT}/M_Z)$, and using
$M_{GUT}=2\times 10^{16}$GeV, $\alpha_3=0.11,0.12,0.13$, we find 
\begin{eqnarray}
\lambda(M_Z) & = & 1.5\ \lambda(M_{GUT}) \nonumber \\
\lambda^{\prime}(M_Z) & = & 3.4,\ 3.6,\ 3.7\ 
\lambda^{\prime}(M_{GUT}) \label{LEHE} \\
\lambda^{\prime\prime}(M_Z) & = & 4.0,\ 4.3,\ 4.7\ 
\lambda^{\prime\prime}(M_{GUT}) \nonumber
\end{eqnarray}
These results break down at the $10\%$ level only when  $\lambda(M_Z)$,
$\lambda^{\prime}(M_Z)$, $\lambda^{\prime\prime}(M_Z)$ exceed 0.27,
0.15, 0.15, which are very much larger than the values which we shall
consider in our FCNC and neutrino calculations. For the R--parity violating
couplings which couple directly to $h_t$ (and, for sufficiently large
$\tan\beta$, to $h_b$) the ratio of the low energy to high energy value
is reduced somewhat, typically by up to around 20\%.

Another interesting feature, and one which has been discussed in a
number of recent papers \cite{br,gs,herbi,roger}, is that it is
possible to derive triviality bounds on the R--parity violating
couplings in exactly the same way as for the top quark Yukawa. Such
bounds depend quite strongly on the values of $\tan\beta$ and
$\alpha_3$. We give bounds derived from our RGEs in Table 1 by simply
setting the coupling to be very large at the GUT scale, either with
small $h_t$ or $h_t$ also at triviality, and running down to a scale of
order the top mass. These results agree quite well with those of
references \cite{br,gs}, and agree also with those given in reference
\cite{roger} where applicable.
\begin{center}
\begin{tabular}{|c|c|c|c|}\hline
Coupling & \ Limit, $\alpha_3=0.11\ $ & \ Limit, $\alpha_3=0.12\ $
& \  Limit, $\alpha_3=0.13\ $ \\ \hline
$\lambda{}$ &   0.92       &  0.92       & 0.92         \\ \hline
$\lp{}$     &   1.08       &  1.10       & 1.12         \\ \hline
$\lpp{}$    &   1.17       &  1.20       & 1.23         \\ \hline
$\l{}$      &  0.92 (1.08) & 0.92 (1.11) & 0.92 (1.13)  \\ \hline
$\lp{xxx}$  &  1.08 (1.08) & 1.10 (1.11) & 1.12 (1.13)  \\ \hline
$\lp{x3x}$  &  1.00 (1.00) & 1.02 (1.03) & 1.04 (1.04)  \\ \hline
$\lpp{xxx}$ &  1.17 (1.08) & 1.20 (1.11) & 1.23 (1.13)  \\ \hline
$\lpp{3xx}$ &  1.02 (0.93) & 1.05 (0.95) & 1.08 (0.96)  \\ \hline
\end {tabular}
\vskip .5cm
\footnotesize{
Table 1. Triviality limits on various couplings derived firstly by 
considering $h_t$ small, and secondly in the case where $h_t$ 
approaches its triviality limit. In the latter case we show the low energy 
value of $h_t$ in brackets. Here ``x'' can represent either 1 or 2 
(or 3 if the index corresponds to an $L$ or $e$ and we are not in the 
very large $\tan\beta$ regime), and the calculation has been performed
for a variety of different $\alpha_3(M_Z)$.}
\end{center}

\section{Choice of Basis}
We shall be interested in deriving constraints on R--parity violation
which arise through lepton flavour violation (LFV), in which processes
are allowed to violate lepton number generation by generation, quark
flavour violation (QFV), or the mixing between leptonic and higgs
superfields (LH mixing). Given the nature of these effects, it will be
convenient here to discuss our choice of basis, and how this basis
affects our calculations. Indeed, it has recently been pointed out that
bounds derived in a quark mass eigenstate basis can be rather different
from those derived with the same couplings in a weak eigenstate basis
\cite{ag}. While the physics is invariant, it is important that we
choose a convenient (and consistent) basis to work in.

It is well known that, by judicious rotations of the fermion fields in
the Standard Model, it is possible to work in a basis where the mass
eigenstates are as far as possible  weak eigenstates, thus eliminating
FCNC at tree level and ensuring lepton flavour conservation at all
orders even through charged current processes. In terms of
supersymmetric parameters, the three fermion mass matrices correspond
to Yukawa matrices $h^e_{ij}$, $h^u_{ij}$, $h^d_{ij}$ of which the
first two can be simultaneously diagonalised. Given the usual
assumptions of universality at the GUT scale, the soft mass matrices
are diagonal while the trilinear terms are proportional to the
corresponding Yukawas, and so there is no LFV, and the only QFV effects
beyond the usual SM ones are those coming from the CKM matrix, which
are relatively small \cite{hagelin,otherfcnc}. This picture can break
down when non-universal soft masses are allowed at the GUT scale, for
example because one is considering a realistic unified theory, but this
is outside the scope of our analysis.

It is clear from the RGEs presented in Appendix A that for some
arbitrary choice of R-parity violating couplings there will be LFV,
QFV, and LH mixing effects. The last case, for example, is driven
through diagrams of the type shown in Figure 1. These generate
non-diagonal anomalous dimension matrix elements of form
$\gamma_{L_i}^{H_1}$ which violate lepton number, and analagous
diagrams exist which conserve $L$ and $B$ but generate LFV or QFV.
With the insertion of appropriate spurion terms for the soft masses 
we will also find LFV and QFV effects in the soft sector.

Thus, even working in a basis where the Yukawa matrices which generate 
the fermion masses are diagonal at the GUT scale, this will not
necessarily be true at the electroweak scale. It is simplest then to
invoke the GIM mechanism and to perform a field rotation so as to
eliminate the most awkward effects and simplify calculations, which we
find to be the quark and lepton mass eigenstate basis. The question now
is what effect this rotation will have on the LFV, QFV and LH mixing
terms in the soft sector.

Continuing with the case of LFV, in general this field rotation must
take the form $L\to UL$, $e\to V^*e$, suppressing all indices for 
simplicity, so that
\begin{equation}
Lh^eeH_1 \to L(U^Th^eV^*)eH_1
\end{equation}
with $U$ and $V$ unitary matrices selected so that $(U^Th^eV^*)$ is
diagonal. We now consider the effect of this rotation on the slepton
mass matrices. We find that
\begin{equation}
m^2_L\to \tilde m^2_L:=U^{\dagger}m^2_LU
\end{equation}
If we assume that 
\begin{equation}
m^2_L=m^2 \times I+\delta m^2
\label{diagform}
\end{equation}
where $I$ is the identity matrix and $\delta m^2$ is a matrix with the
only non-zero elements small and off-diagonal, while similarly
$U=I+O(\varepsilon)$ with $\varepsilon$ parametrising the rotation and
assumed to be small, we can see that
\begin{equation}
\tilde m^2_L=m^2_L + O(\varepsilon\delta m^2)
\end{equation}
Clearly, if we have sufficiently small off-diagonal elements in the
mass and Yukawa matrices, we are quite safe in taking the non-diagonal
terms in the mass-matrix to be the same before and after we have
rotated back into the appropriate mass eigenstate basis. It is
straightforward to check that this argument is valid for all the LFV
effects which we consider.

The argument above assumes degenerate masses for the relevant
generations and so will clearly be inaccurate in the case of any
process involving the third generation, such as $b\to s\gamma$, and for
the case where we consider $LH$ mixing. Here there is no alternative
but to perform the rotation into the appropriate basis explicitly. The
case for $LH$ mixing is somewhat different, since here the preferred
basis for simplicity is not a mass eigenstate basis but one in which
there are no couplings in the superpotential of form $\mu_iL_iH_2$, but
the same principle applies. As will be shown explicitly in the next
section, even continuously rotating the fields so as to ensure that
$\mu_i$ remains zero at all scales will not remove all the $LH$ mixing
effects in, for example, the running of the Yukawa couplings, the soft
bilinear terms, and $m^2_{L_iH_1}$.

We conclude by noting that in the fermion mass eigenstate basis the
non-diagonal soft masses will appear in the form of flavour violating
mass insertions in our diagrams, rather than in the vertices
themselves.

\section{Sneutrino VEVs}
\subsection{General Discussion}
One signature of R-parity violation which has been much discussed is
that of sneutrino VEVs. These typically arise because of the existence
of $\mu_i$, $D_i$, and $m^2_{L_iH_1}$ terms which explicitly cause the
effective potential to contain terms linear in the sneutrino field,
either from explicit \cite{suzuki,lee} or spontaneous \cite{rv}
R-parity violation (the origin of R-parity violation is irrelevant if
it appears in this sector, and recently the implications of such
effects have been considered in the context of GUTs \cite{vissani}).
They can also be caused by one loop effects involving dimensionless
R--parity violating couplings \cite{bgmt,enqvist}.

Once sneutrinos have acquired VEVs, neutrinos and neutralinos
mix\footnote{As in fact, charginos and leptons, and Higgses and
sneutrinos}, so that we may derive bounds on R-parity violating terms
by imposing experimental limits on neutrino masses. In fact, it has
recently been shown that it is possible not merely to bound R--parity
violating terms, but to select such terms so as to explain the present
rather complicated array of experimental measurements of neutrino
masses and oscillations \cite{hemp}. 

The potential which we must minimise takes the form
\begin{eqnarray}
V&=&(m_{H_1H_1}^2+\mu_4^2)\nu_1^2+(m_{H_2H_2}^2+\mu_4^2)\nu_2^2
+2D_4\nu_1\nu_2
+\frac{g_1^2+g_2^2}{8}(\nu_1^2-\nu_2^2)^2 \nonumber  \\
&&+2(m_{H_1L_i}^2+\mu_4\mu_i)\nu_1l_i+2D_il_i\nu_2 \\
&&+(m_{L_iL_j}^2+\mu_i\mu_j)l_il_j
+\frac{g_1^2+g_2^2}{4}l_i^2(\nu_1^2-\nu_2^2)
\nonumber
\end{eqnarray}
Here $l_i$ is the VEV of $L_i$, with $i$, $j$ running from 1 to 3, we
drop all terms of order $l^3$ and above (since we expect $l\ll \nu$).
We define as usual $\nu_i=<H_i>$, with
$\nu^2=\nu_1^2+\nu_2^2=(174\hbox{GeV})^2$ and $\tan\beta=\nu_2/\nu_1$.

In order to study these effects, it is most convenient to follow
reference \cite{suzuki}, and perform a field rotation mixing the $L_i$
and $H_1$ such that only one of the $\mu_i$, which we
choose to be $\mu_4$, is non-zero. Doing so will not merely change the
dimensionful couplings, but will also generate new contributions to the
dimensionless couplings $\lambda$ and $\lambda^{\prime}$ from $h_\tau$
and $h_t$. In this basis, however, both the Higgs-sneutrino potential
and the neutrino-neutralino mass matrix take on relatively simple
forms. Up to order $l/\nu$, and assuming that the off-diagonal terms in
the mass matrix $m^2_{L_iL_j}$ are small, we find that
\begin{equation}
l_i=-
\frac{D_i\nu_2+m^2_{H_1L_i}\nu_1}{m^2_{L_i}+\frac{1}{2}M_Z^2\cos 2\beta}
\label{sVEV}
\end{equation}
There are two dangerous approximations which we have made and which
should be justified. The first is that we have only considered the
Higgs potential at tree level. This is well known not to be reliable in
general, but we would not expect that the large radiative corrections
due to quark and squark loops should greatly affect the calculation of
sneutrino VEVs. Hence we expect our analysis to be reliable so long as
we derive the Higgs VEVs $\nu_1$, $\nu_2$ from the full one-loop
effective potential before substituting them into
equation~(\ref{sVEV}).

Our second approximation is that we have not considered the processes
discussed in detail in reference \cite{enqvist}, in which one-loop
diagrams generate sneutrino VEVs even with all dimensionful R-parity
violating couplings zero. We shall ignore this point, and simply draw
the reader's attention to the fact that the effects we study will in
general be supplementary to these.

The neutralino-neutrino mass matrix then takes the form, in the basis
of 2--spinors given by
$\{i\tilde{B^0},i\tilde{W_3^0},\tilde{H_1^0},\tilde{H_2^0}, L_i\}$,
\begin{equation}
\left( \begin{array}{ccccc}
M_1 & 0 & -\frac{1}{\sqrt{2}}g_1\nu_1 &
     \frac{1}{\sqrt{2}}g_1\nu_2 & -\frac{1}{\sqrt{2}}g_1l_i \\
0 & M_2 & {\frac{1}{\sqrt{2}}}g_2\nu_1 & 
   -\frac{1}{\sqrt{2}}g_2\nu_2 & \frac{1}{\sqrt{2}}g_2l_i \\
-\frac{1}{\sqrt{2}}g_1\nu_1 & \frac{1}{\sqrt{2}}g_2\nu_1 &
      0 & \mu_4 & 0 \\
\frac{1}{\sqrt{2}}g_1\nu_2 & -\frac{1}{\sqrt{2}} g_2\nu_2 &
     \mu_4 & 0 & 0 \\ 
-\frac{1}{\sqrt{2}}g_1l_i & \frac{1}{\sqrt{2}}g_2l_i & 0 & 0 & 0
\end{array} \right)
\label{mmatrix}
\end{equation}
suggesting that the neutrino mass generated is of order
$(g_1^2+g_2^2)l_i^2/2M$, where $M$ is some typical neutralino mass.

The simplest constraints on neutrino masses for the first generation
are the direct bounds as quoted for example in reference \cite{pdb}. We
could also use experimental bounds on neutrinoless double beta decay,
on neutrino oscillation, and from cosmology. The oscillation effects
are rather complicated and we shall not consider them in this analysis
except to note that the latter would give constraints on two or more
R-party violating couplings considered together, similarly to the case
where the neutrino masses and mixings are generated directly from the
dimensionless couplings, and that in any case products of $\l{}$s are
already very tightly constrained by these effects
\cite{bgmt,enqvist,bm}. In light of the complexity of our results, we
shall not derive bounds as such, but instead will discover how large
the typical impact of the R--parity violating couplings can be.

\subsection{Generation of Sneutrino VEVs from the RGEs}
Given the discussion of sneutrino VEVs above, we wish to investigate
whether the RGEs can tell us anything about the likely size of such
effects. The usual scenario has supersymmetry breaking in the hidden
sector communicated to the visible sector only by gravity, so that the
soft breaking terms take the form of trilinear and bilinear terms
proportional to the corresponding Yukawa couplings and $\mu$ terms,
with universal soft masses. If we make this assumption, then at the GUT
scale (or perhaps more plausibly the Planck scale) we can simply rotate
away the $\mu_i$ for $i\ne 4$ as described above, and in doing so we
will also set all $D_i=0$ except for $D_4$ without generating any
non-zero $m^2_{H_1L_i}$. Hence we shall assume that at the GUT scale we
have done this, and so our initial configuration will have R-parity
violation only through dimensionless couplings and trilinear terms.

If we then run our initial conditions down to the weak scale, then we
will find that here sneutrino VEVs will only be generated through the
following non-zero products of couplings.
\begin{equation}
\begin{array}{cccl}
\l{i33}h_\tau  & \hbox{or} & \lp{i33}h_b  &
    \hbox{generating $\mu_i$, $D_i$, $m^2_{H_1L_i}$} \\
\C{i33}h_\tau  & \hbox{or} & \Cp{i33}h_b  & 
    \hbox{generating $D_i$} \\
\C{i33}\eta_\tau & \hbox{or} & \Cp{i33}\eta_b  &
    \hbox{generating $m^2_{H_1L_i}$}
\end{array}
\end{equation}
These of course are exactly what we would expect, since in order to
generate sneutrino VEVs we must have terms mixing $L_i$ and $H_1$,
as shown in Figure 1.

Once we run down to the weak scale, however, we must perform another
change of basis to that in which $\mu_i=0$, and this will introduce
additional effects. The simplest way to describe these is by recasting
the relevant RGEs in the basis where we continuously rotate as we run so as
to ensure that 
\begin{equation}
\frac{d\mu_i}{dt}= 0
\end{equation}
In this basis we find that for $i\ne 4$ and assuming that the diagonal
terms in the soft mass matrices are very much greater than the off-diagonal
terms and $\mu_4\gg\mu_i$
\begin{eqnarray}
16\pi^2 \frac{dD_i}{dt}&\simeq& 
      D_i(3h_t^2 - g_1^2 - 3g_2^2)
    - 6\Cp{i33}h_b\mu_4 - 2\C{i33}h_\tau\mu_4 \nonumber \\
16\pi^2 \frac{d m^2_{H_1L_i}}{dt}&\simeq& 
         - \l{i33}h_\tau (2m^2_{L_i}+2m^2_{e_3}+2m^2_{L_3})
\label{Dmeqn} \\
      && - \lp{i33}h_b (6m^2_{L_i}+6m^2_{d_3}+6m^2_{Q_3}) \nonumber \\ 
      && - 2\C{i33}\eta_\tau - 6\Cp{i33}\eta_b
\nonumber
\end{eqnarray}
Note that although it is possible to cancel off some of the $L_i$-$H_1$
mixing terms by a field redefinition, not all terms can be cancelled
simultaneously. Here for example the rotation has removed the $m_{H_1}$
terms from the RGE for $m^2_{H_1L_i}$, but increased the coefficient of
the $m_{L_i}$ terms.

We see that the effects are likely to be largest when $\tan\beta$ is
large, giving large $h_b$ and $h_\tau$, but that the dependence is
clearly rather complicated. More significantly, the sneutrino VEV will
in general be proportional to the R-parity violating coupling, and
hence the neutrino mass to the coupling squared.

\subsection{Numerical Results}
Many of the numerical methods and assumptions which we use here will be
standard for calculations throughout this paper. We have as input
parameters $m_t$, $\alpha_3$, $m_0$, $A_0$, $M_{1/2}$,
$\tan\beta=\nu_2/\nu_1$ and the sign of $\mu$. We run the Yukawa and
gauge couplings up to the GUT scale, where we impose universality of
the soft masses, but not unification, and set any R-parity violating
couplings which we wish to investigate. $m_0^2$ and $M_{1/2}$ are the
values of all the diagonal elements of the scalar soft mass-squared
matrices and the universal gaugino mass respectively. We set each
trilinear coupling equal to the corresponding Yukawa coupling
multiplied by $A_0$, so that for example
$\Cp{ijk}(M_{GUT})=\lp{ijk}(M_{GUT})A_0$. The masses and couplings are
then run down to low energy to give output. $\mu_4$ and $D_4$ are
chosen so as to give the correct minimum of the one-loop effective
potential. We do not include any threshold corrections at either scale,
using the non-SUSY RGEs below $M_{1/2}$ and the SUSY RGEs above it.

Typical results are shown in Figure 2a, where we set
$\l{133}(M_{GUT})=0.01$ and Figure 2b where instead we set
$\lp{133}(M_{GUT})=0.001$. Input parameters are  $m_t=175$GeV,
$\alpha_3(M_Z)=0.12$, $A_0=0$, $M_{1/2}=500$GeV, and we plot the
absolute magnitude of the sneutrino VEV and the corresponding neutrino
mass as a function of $m_{L_1}$, the soft mass of the left-handed
slepton, for both signs of $\mu$ and for $\tan\beta=$2,20. We note that
in certain cases the sneutrino VEV changes sign at some value where the
contributions for the non-zero $D_1$ and $m^2_{H_1L_1}$ exactly cancel.
It is clear from equation~(\ref{Dmeqn}) that, for given $A_0$ and for
one or other sign of $\mu_4$, $D_1$ and $m^2_{H_1L_1}$ will have
opposite sign, and here we see that complete cancellation can occur
when $\mu_4>0$. Here larger $\tan\beta$ gives a larger sneutrino VEV
and hence larger neutrino mass.

Although we might na{\"\i}vely expect from the form of equation
(\ref{mmatrix}) that increasing $M_{1/2}$ (for fixed $m_{L_i}$) should
cause the resulting neutrino mass to decrease, increasing $M_{1/2}$
also changes the value of $\mu_4$ and affects the running of all the
dimensionful parameters and so the situation is rather complicated. In 
Figure 3 we plot the absolute magnitude of the sneutrino VEV and the
resulting neutrino mass against $M_{1/2}$ for the same parameters as
Figure 2b, but with $m_{L_1}$ held fixed at 500GeV. Again we see that
the relative cancellation occurs to give zero neutrino mass for some
set of input parameters, and is noticable that even with $M_{1/2}$ as
large as 500GeV, the neutrino mass is still increasing with
increasing $M_{1/2}$ in at least some cases.

The effect of varying $A_0$ is also large, as shown in Figure 4. The
main limit on the size of $A_0$ is that its absolute value should be
less than around 3 times a typical slepton mass to avoid charge or
colour breaking minima \cite{ds}. In Figure 4 we vary $A_0$ over the
whole of this range for the same parameters as in Figure 2b, but with
$m_{L_i}=M_{1/2}=500$GeV. From this figure it is clear that the
cancellation of the effect can occur for either sign of $\mu$,
depending on the signs and relative magnitudes of the low energy value
of the trilinear couplings.

We now turn to a discussion of how these calculations can be used to
derive a limit on the couplings. It is clearly straightforward to
derive bounds on $\l{i33}$ and $\lp{i33}$ for a point on one or other
of Figures 2a to 4 by calculating which value of $\l{}$ or $\lp{}$
gives a neutrino mass which is acceptable in the light of the
experimental limits, given that the sneutrino VEV is directly
proportional to $\l{133}$ (or $\lp{133}$), and so the neutrino mass
generated is proportional to the square of each of the coupling. What
is perhaps most interesting here is that although the constraint tends
to become less tight with increasing values of the soft masses, as is
generally the case for the constraints on R-parity violating couplings,
the decrease can be slow relative to the case for limits derived from
one-loop diagrams with internal sfermion lines, as these are almost
always power-law suppressed by soft masses. We typically find that even
for soft masses of order 500GeV the bounds are relatively tight, with
$\l{i33}(M_{GUT})\sim 0.01$ and $\lp{i33}(M_{GUT})\sim 0.001$ both
giving neutrino masses of order 1keV or more, suggesting a bound on
these couplings which is at least an order of magnitude tighter for the
first generation. These constraints are much tighter than those in the
literature on individual couplings \cite{bgh,bpw,bgnn}, since normally
the tightest constraints are on products of couplings. It should however
be noted that we are quoting limits on couplings at the GUT scale,
which should be converted to values at the weak scale using equation
(\ref{LEHE}) for comparison with those given by most other authors.

Such limits are obviously greatly hampered by the dependence on the
many input parameters, on the assumptions about the GUT scale
structure, and in particular by the fact that there are strong
cancellations between terms. We note that one possible cause of partial
cancellation, namely having $\l{i33}$ and $\lp{i33}$ of the same order
and opposite sign, is unlikely in the context of certain GUTs, where we
would typically expect $\l{i33}(M_{GUT})=\lp{i33}(M_{GUT})$ just as
$h_b(M_{GUT})=h_\tau(M_{GUT})$. Hence our results should be regarded as
indicative of the likely magnitude of the effect rather than being able
to strongly constrain the electroweak scale values of the couplings in
isolation. Nevertheless, it is significant that the RGEs typically give
such large effects, and in constructing any model which includes any of
the couplings which can generate neutrino masses it is necessary either
to set them extremely small or else to fine-tune away the unwanted
effects.

\section{$\mu\to e\gamma$}
\subsection{Contributions}
The process $\mu\to e\gamma$ is one of the most tightly constrained
examples of FCNC. Since this process cannot occur in the SM, its
observation would be pressing evidence for new physics beyond the SM.
Experiments \cite{megex} have calculated an upper bound on the
branching ratio (BR) of 
\begin{equation}
BR(\mu\to e\gamma)< 4.9\times 10^{-11}\hbox{ at 90\% CL}
\end{equation}
In SUSY models, a non-zero rate can be generated through non-diagonal
slepton mass matrices \cite{hagelin,megsusy,us}, and also through the
direct effects of R--parity violating couplings \cite{suzuki,lee,bgmt}.
However, as noted above, R--parity violation induces LFV through soft
terms, and so in any realistic model where the latter effects occur, so
will the former, and here we shall consider the two effects together.

The total branching ratio for the process $\mu\to e\gamma$ can be written as:
\begin{equation}
BR (\mu\to e\gamma) = \frac{12 {\pi}^2}{G_F^2} \left(|\tilde{A}_{LR}|^2 + 
|\tilde{A}_{RL}|^2 \right)
\label{br}
\end{equation}
where $G_F$ is the Fermi constant, and $\tilde{A}_{LR}$,
$\tilde{A}_{RL}$  are the total amplitudes to the LR and RL
transitions respectively. Here
\begin{equation}
\tilde{A}_i = \tilde{A}_i^{\lambda} + 
\tilde{A}_i^{\lp{}} + \tilde{A}_i^{\Delta m}
\end{equation}
with $i=LR,RL$ and the different terms are as follows. The various
contributions to $\tilde A_{LR}^{\lambda}$, $\tilde A_{LR}^{\lp{}}$ and
$\tilde A_{LR}^{\Delta m}$ are shown in Figures 5 to 7 respectively.
Diagrams contributing to $\tilde A_{RL}$ are not shown, but are similar
to the $LR$ case for $\tilde A_{RL}^{\lambda}$ and the neutralino
contributions to $\tilde A_{RL}^{\Delta m}$. Both  $\tilde
A_{RL}^{\lp{}}$ and the chargino contributions to $\tilde
A_{RL}^{\Delta m}$ are proportional to the electron mass and thus
neglected.

Results for the various amplitudes are :
\begin{equation}
\tilde{A}_{LR}^{\lambda}  =  \frac{e}{16 \pi^2} \sum_{i,j=1}^{3}
\l{i1j} \l{i2j} \frac{1}{6} \left( 
\frac{1}{m^2_{\tilde\nu_i}} - \frac{1}{2} \left[ 
\frac{\sin^2 \theta_{e_j}}{m^2_{\tilde{e}_j^{(1)}}} 
+ \frac{\cos^2 \theta_{e_j}}{m^2_{\tilde{e}_j^{(2)}}} \right] \right) \; ,
\label{allr}
\end{equation}
where $e$ is the electromagnetic constant, $m_{\tilde{\nu}_i}$ is the 
mass of the corresponding sneutrino, $m_{\tilde{e}_j^{(1),(2)}}$
denote the two mass eigenvalues of the mass matrix for the $j^{th}$ 
generation of sleptons and $\sin\theta_{e_j}$, $\cos\theta_{e_j}$
are elements of the corresponding orthogonal diagonalisation matrix. 
These are given by:
\begin{eqnarray}
\sin^2 \theta_{e_j} & = & \frac{1}{2} \left( 1-
\frac{m^2_{\tilde{e}_j^L} - m^2_{\tilde{e}_j^R}}{m^2_{\tilde{e}_j^{(1)}}
-m^2_{\tilde{e}_j^{(2)}}} \right) \nonumber \\
\cos^2 \theta_{e_j} & = & \frac{1}{2} \left( 1+
\frac{m^2_{\tilde{e}_j^L} - m^2_{\tilde{e}_j^R}}{m^2_{\tilde{e}_j^{(1)}}
-m^2_{\tilde{e}_j^{(2)}}} \right)
\label{thetas}
\end{eqnarray}
so that state $e^{(1)}$ is defined by
\begin{equation}
\vert e^{(1)}> = \cos\theta_{e}\vert e_L> + \sin\theta_{e}\vert e_R>
\end{equation}

The second amplitude is given by: 
\begin{eqnarray}
\tilde{A}_{LR}^{\lp{}} & = & \frac{e}{16 \pi^2} \sum_{i,j=1}^{3}
\lp{1ij} \lp{2ij} \left( 
\frac{\cos^2 \theta_{u_i}}{m^2_{\tilde{u}^{(1)}_i}} 
(F_1(x^{(1)}_{ji})+2F_2(x^{(1)}_{ji})) \right. \nonumber \\
& + &  \left.
\frac{\sin^2 \theta_{u_i}}{m^2_{\tilde{u}^{(2)}_i}} 
(F_1(x^{(2)}_{ji})+2F_2(x^{(2)}_{ji})) - 
\frac{\sin^2 \theta_{d_j}}{m^2_{\tilde{d}^{(1)}_j}}
(2F_1(x'^{(1)}_{ij})+F_2(x'^{(1)}_{ij})) \right. \label{alplr} \\
& - &  \left. \frac{\cos^2 \theta_{d_j}}{m^2_{\tilde{d}^{(2)}_j}}
(2F_1(x'^{(2)}_{ij})+F_2(x'^{(2)}_{ij})) \right)  \nonumber \; .
\end{eqnarray}
The various functions are mostly found in \cite{us} and \cite{bert},
and are reproduced in Appendix B for convenience, while
$x^{(1),(2)}_{ji}=\frac{m^2_{d_j}}{m^2_{\tilde{u}^{(1),(2)}_i}}$,
$x'^{(1),(2)}_{ij} = \frac{m^2_{u_i}}{m^2_{\tilde{d}^{(1),(2)}_j}}$.

Finally the third amplitude is:
\begin{eqnarray}
\tilde{A}_{LR}^{\Delta m} & = & \frac{eg^2}{8\pi^2} \left\{ 
-\frac{1}{2}
\frac{\Delta m^2_{\tilde{\nu}_e \tilde{\nu}_{\mu}}}{m^4_{\tilde{\nu}}}
\sum_{j=1}^{2} \left( |V_{j1}|^2 G(x_j) - 
\frac{V_{j1}U_{j2}}{\sqrt{2}\cos\beta} \frac{M_{\chi_j^-}}{M_W} H(x_j)
\right) \right. \cr
& + & \left. \frac{\Delta m^2_{\tilde{e}_L \tilde{\mu}_L}}{m^4_{\tilde{e}_L}} 
\sum_{j=1}^{4} \left( \left| s_W N'_{j1}+\frac{1}{c_W} \left(
\frac{1}{2} -s_W^2 \right) N'_{j2} \right|^2 
 F(x_{jL})  \right. \right.
\nonumber \\
&& \qquad - \left. \left. \left[ s_W N'_{j1} + \frac{1}{c_W} \bigl( 
\frac{1}{2}-s_W^2 \bigr ) N'_{j2} \right]
\frac{N_{j3}}{2 \cos \beta} \frac{M_{\chi^0_j}}{M_W} L(x_{jL}) \right)
\right. \label{amlr} \\
& + & \left. \frac{\Delta m^2_{\tilde{e}_L 
\tilde{\mu}_L}}{m^2_{\tilde{e}_L} - m^2_{\tilde{e}_R}} 
\sum_{j=1}^{4} \left[ s_W N'_{j1} + \frac{1}{c_W} \left( \frac{1}{2}
-s_W^2 \right) N'_{j2} \right] \left( -s_W N'_{j1} + \frac{s_W^2}{c_W}
N'_{j2} \right) \right. \nonumber \\
& \times & \left. M_{\chi^0_j} (A_{\mu}+\mu \tan\beta) \left( 
\frac{1}{m^2_{\tilde{e}_L}-m^2_{\tilde{e}_R}} \left[
\frac{F_4(x_{jL})}{m^2_{\tilde{e}_L}} -
\frac{F_4(x_{jR})}{m^2_{\tilde{e}_R}} \right] +
\frac{L(x_{jL})}{m^4_{\tilde{e}_L}} \right) 
\right\} \nonumber \; \;,
\end{eqnarray}
where $s_W$ ($c_W$) is the sine (cosine) of the Weinberg angle, and
$g$ is the SU(2) gauge coupling constant. Also,
\begin{equation}
m^2_{\tilde{\nu}_e}=m^2_{\tilde{\nu}_{\mu}} \equiv m^2_{\tilde{\nu}},
\qquad
m^2_{\tilde{e}_L}=m^2_{\tilde{\mu}_L},
\qquad
m^2_{\tilde{e}_R}=m^2_{\tilde{\mu}_R} \;,
\end{equation}
and 
\begin{equation}
x_j=\frac{M^2_{\chi_j^-}}{m_{\tilde\nu}^2}, \quad
x_{jL}=\frac{M^2_{\chi_j^0}}{m_{\tilde e_L}^2}, \quad
x_{jR}=\frac{M^2_{\chi_j^0}}{m_{\tilde e_R}^2}
\end{equation}
and the mixing matrices $U$, $V$, $N$, $N'$ are defined as in
\cite{gunhab}. Finally $A_{\mu}$ is the trilinear coupling for the
second generation of sleptons. We use $\Delta m^2_{\tilde{\nu}_e 
\tilde{\nu}_{\mu}} =\Delta m^2_{\tilde{e}_L\tilde{\mu}_L}$ and
$\Delta m^2_{\tilde{e}_R\tilde{\mu}_R}$ instead of $m^2_{L_1L_2}$
and $m^2_{e_1e_2}$ as in Appendix A to avoid confusion between
generation and mass eigenstate labels.

The RL amplitudes are given by:
\begin{equation}
\tilde{A}_{RL}^{\lambda}  =  \frac{e}{16 \pi^2} \sum_{i,j=1}^{3}
\l{ij1} \l{ij2} \frac{1}{6} \left(
\frac{1}{m^2_{\tilde\nu_i}} - \frac{1}{2} \left[
\frac{\cos^2 \theta_{e_j}}{m^2_{\tilde{e}_j^{(1)}}}
+ \frac{\sin^2 \theta_{e_j}}{m^2_{\tilde{e}_j^{(2)}}} \right] \right) \; ,
\label{alrl}
\end{equation}
with $\tilde{A}_{RL}^{\lp{}}=0$ and:
\begin{eqnarray}
\tilde{A}_{RL}^{\Delta m} & = & \frac{eg^2}{8\pi^2} \left\{
\frac{\Delta m^2_{\tilde{e}_R \tilde{\mu}_R}}{m^4_{\tilde{e}_R}}
\sum_{j=1}^{4} \left( \left| -s_W N'_{j1}+\frac{s_W^2}{c_W} N'_{j2} 
\right|^2 F(x_{jR})  \right. \right.\nonumber \\
&&\qquad - \left. \left. \left[ -s_W N'^{*}_{j1} + \frac{s_W^2}{c_W} 
N'^{*}_{j2} \right] \frac{N'^{*}_{j3}}{2 \cos \beta} 
\frac{M_{\chi^0_j}}{M_W} L(x_{jR}) \right) \right. \label{amrl} \\
& + & \left. \frac{\Delta m^2_{\tilde{e}_R
\tilde{\mu}_R}}{m^2_{\tilde{e}_R} - m^2_{\tilde{e}_L}}
\sum_{j=1}^{4} \left( -s_W N'^{*}_{j1} + \frac{s_W^2}{c_W} 
N'^{*}_{j2} \right) \left( s_W N'^{*}_{j1} + \frac{1}{c_W} \left(
\frac{1}{2}-s_W^2 \right) N'^{*}_{j2} \right) \right. \nonumber \\
& \times & \left. M_{\chi^0_j} (A_{\mu}+\mu \tan\beta) \left(
\frac{1}{m^2_{\tilde{e}_R}-m^2_{\tilde{e}_L}} \left[
\frac{F_4(x_{jR})}{m^2_{\tilde{e}_R}} -
\frac{F_4(x_{jL})}{m^2_{\tilde{e}_L}} \right] +
\frac{L(x_{jR})}{m^4_{\tilde{e}_R}} \right)
\right\} \nonumber
\end{eqnarray}
In calculating these amplitudes, we have adopted the
usual approximation that the electron mass is negligible relative to
the muon mass, and we work to first order in both $m_\mu/M_W$ and
$m_\mu(A_\mu+\mu\tan\beta)/(m_{\tilde e_L}^2-m_{\tilde e_R}^2)$.

Finally we note that we have neglected many diagrams which are
suppressed by the mixing between neutralinos and neutrinos, or between
charginos and charged leptons \cite{suzuki,lee}. Two typical examples of
such diagrams contributing to $\tilde A_{LR}$ are shown in Figure~8.
Such terms will exist due to mixing of neutral fermions as shown by the
mass matrix of equation (\ref{mmatrix}), and similarly from the mass
matrix for charginos and sleptons which contributes to the lagrangian
(in two-spinor notation)
\begin{equation}
(-i\tilde{W^-},\tilde H_1^-,e_{Li} )
\left(
\begin{array}{ccc}
 M_2 & g_2\nu_2 & 0 \\
 g_2\nu_1 & \mu_4 & h_il_i \\
 g_2l_i & 0 & h_i\nu_1 \\
\end{array} \right)
\left( \begin{array}{c}
 -i\tilde{W^+} \\
 \tilde H_2^+ \\
 e^c_{Ri} \\
\end{array} \right)
\end{equation}
Here we have presented only one generation for simplicity, and $h_i$ is
its mass-generating Yukawa coupling. However, even for the third
generation the sneutrino VEV $l_i$ must be quite small to avoid an
excessive neutrino mass, and hence the mixing is small. Relative to the
direct $\lambda^2$ proportional contributions shown in Figure 5 such
diagrams with a predominantly lepton line are suppressed by
$g_2^2l_i/(\lambda M)$ (where a factor $g_2/\lambda$ is from the
differing couplings, and $g_2l_i/M$ is from  the mixing angle). Thus we
expect this diagram to contribute less than the direct contribution
except when $\lambda$ is so small that both are negligible, in which
case we  would anyway not expect the generation of large $l_i$, since
from our earlier discussion of sneutrino VEVs we always have
$l_i<\l{}M$ by at least an order of magnitude. Similarly, if  the
internal fermion line is predominantly neutralino or chargino, such
contributions are similar to those shown in Figure 7 but suppressed by
a relative factor $l_i\lambda/M$, and so negligible.

Further diagrams can be drawn with mixing between sneutrinos and
neutral Higgs states, and between charged sleptons and charged Higgs
states, but by similar arguments we are justified in neglecting those
too. We note however that although we expect such diagrams to give
weaker bounds that for the case where corresponding diagrams without
sneutrino VEV insertions exist, they can provide very weak constraints
on products of couplings which we cannot otherwise bound. For example
Figure~8a will allow a bound on the product $\l{i33}\l{12i}$.

\subsection{Analytical Discussion}
Given the complexity if the expressions shown above, we can try to see
whether in certain limits we are able to derive analytical bounds on
the R--parity violating couplings. In each case, we shall set only two
R-parity violating dimensionless couplings non-zero at $M_{GUT}$ and
see what effects they generate. We begin by discussing the relatively
simple ``direct'' contributions to $\tilde A^{\lambda}$ and $\tilde
A^{\lp{}}$ before turning to the ``indirect'' contributions of $\tilde
A^{\Delta m}$.

Due to the antisymmetry of the $\l{ijk}$ couplings
with respect to the first two indices, we will always have $i=3$ in 
equation~(\ref{allr}) and therefore
$m_{\tilde{\nu}_i}=m_{\tilde{\nu_{\tau}}}$. Also note that, for the
first two generations, the weak interaction and the mass eigenstates
are practically the same, so we can take then $\sin^2
\theta_{e_k}=0$, $\cos^2 \theta_{e_k}=1$ and also
$m^2_{\tilde{e}_k^{(1),(2)}}=m^2_{\tilde{e}_k^{L,R}}$, with $k=1,2$.
This gives us a bound on products of $\lambda$s at the low--energy
scale
\begin{equation}
|\l{31k}(M_Z) \l{32k} (M_Z)| < 4.6\times 10^{-4} \left(
 \frac{m}{100\hbox{Gev}} 
\right)^2 \; \;\;\; k=1,2,
\label{bl12}
\end{equation}
assuming that $m_{\tilde\nu_{\tau}}\simeq m_{\tilde{e}_R}\simeq m$.
Bounds for the case $k=3$ are pretty similar to these, though their
dependence on the mixing in the stau sector makes them less
straightforward. However in the case of maximal mixing (that
is, $\sin^2 \theta_{\tau} = \cos^2 \theta_{\tau} = \frac{1}{2}$, and
with one stau very much lighter than the other) we can
see that:
\begin{equation}
|\l{313}(M_Z) \l{323}(M_Z)| < 2.3\times 10^{-4} \left[ \left(
\frac{100\hbox{Gev}}{m_{\tilde{\nu}_{\tau}}} \right)^2
-  \left(\frac{100\hbox{Gev}}{2m_{\tilde{\tau}_2}} \right)^2
 \right]^{-1}
\; \;,
\label{bl3}
\end{equation}
where $m_{\tilde{\tau}_2}$ is the lighter stau mass eigenvalue.
Therefore we see that there is a possibility of cancellation inside the
bracket, thus considerably reducing the branching ration generated by
this process and hence relaxing the bound. A further problem with these
analytical limits on slepton mediated diagrams is that we are assuming
that the right and left handed sleptons have the same masses, and in
practice this assumption can break down completely, as will become
apparent when we come to consider our exact numerical results.

We turn now to possible limits on these couplings coming from the
RL contributions to $\mu\rightarrow e\gamma$ given in
eq.~(\ref{allr}). In this case we have one pair $\l{ij1}$, $\l{ij2}$,
with $i,j=1,2$, and $i\neq j$ which satisfy the bound 
\begin{equation}
|\l{ij1}(M_Z) \l{ij2} (M_Z)| < 2.3\times 10^{-4} \left(
 \frac{m}{100\hbox{Gev}}
\right)^2 \; \;\;\; i,j=1,2,
\label{blRL}
\end{equation}
Note that there is a relative factor of two between equations
(\ref{bl12}) and (\ref{blRL}) from the fact that we must set
$\l{ij1}=-\l{ji1}$ and so cannot set only one $\l{}\neq 0$ at once. The
remaining cases, $i=3$, $j=1,2$, allow cancellations in a similar
way to the case of equation~(\ref{bl3}).

Concerning the $\lp{}$ couplings, we can turn back to eq.~(\ref{alplr})
and notice that for $i,j=1,2$ again the weak  interaction and mass
eigenstates are the same, so that
$\sin^2\theta_{u_i}=\sin^2\theta_{d_j}=0$,
$\cos^2\theta_{u_i}=\cos^2\theta_{d_j}=1$ and 
$m^2_{\tilde{u}_i^{(1),(2)}}=m^2_{\tilde{u}_i^{L,R}}$,
$m^2_{\tilde{d}_j^{(1),(2)}}=m^2_{\tilde{d}_j^{L,R}}$. Furthermore, due
to the small values of the quark masses for these first two 
generations, $x_{ji}$ and $x'_{ij}$ are very small as well, and 
$F_1(x)$, $F_2(x)$ can be replaced by $1/6$, $1/12$ respectively. All
this allows us to deduce that
\begin{equation}
|\lp{1ij}(M_Z)\lp{2ij}(M_Z)|< 4.6\times 10^{-4} \left(
 \frac{m}{100\hbox{Gev}} 
\right)^2 \; \;\;\; i,j=1,2
\label{blp}
\end{equation}
where here we assume that the squark masses are degenerate at $m$.
However when either $i$, $j$ or both are equal to 3, everything is
going to depend on the mixing between stops and/or sbottoms; in that
case we find a tighter bound (a factor of 2 smaller that for the other
couplings) when that mixing is large by assuming that the contribution
of the lightest mass eigenvalue (sbottom for $i=1,2$ and $j=3$, and
stop for the rest) will be the dominant one in the corresponding
amplitude.
 
So far we have presented analytical bounds on products of couplings at
the electroweak scale. However, note that these can immediately be
translated into bounds on the same couplings at $M_{GUT}$, just by
performing the corresponding running up from $M_Z$. Moreover, the
approximate solutions to the RGEs sketched in eq.(\ref{LEHE}) will
provide the reader with automatic bounds for most of the relevant
cases.

By looking at eqs.~(\ref{amlr}), (\ref{amrl}) it is clear that not much
can be extracted from them analytically except in limits so simple as
not to be useful. However, we can at least remark upon their sign and
order of magnitude. If we consider the corresponding RGEs
for LFV with diagonal mass terms dominant,
\begin{eqnarray}
16\pi^2 \frac{d m^2_{e_ie_j}}{dt} & = &
         \sum_{mn} \Bigl(\l{mni}\l{mnj}
             \bigl(m^2_{e_i} + m^2_{e_j}
                + 4m^2_{L_n} \bigr)\cr
            && \qquad\qquad + 2\C{mni}\C{mnj} \Bigr)
     \\
16\pi^2 \frac{d m^2_{L_iL_j}}{dt} & = &
         \sum_{mn} \Bigl(\l{imn}\l{jmn}
                  \bigl(m^2_{L_i} + m^2_{L_j}
                     + 2m^2_{e_n} + 2m^2_{L_m}\bigr)\cr
            && \qquad\qquad + 2\C{imn}\C{jmn} \Bigr ) \cr
      &+& \sum_{mn}\Bigl(3\lp{imn}\lp{jmn}
                  \bigl(m^2_{L_i} + m^2_{L_j}
                     + 2m^2_{d_n} + 2m^2_{Q_m}\bigr)\cr
            && \qquad\qquad + 6\Cp{imn}\Cp{jmn}\Bigr) \; \; \; ,
\end{eqnarray}
we can solve these equations only in the limit where we only consider
the contributions from the largest gauge coupling after imposing
unification at $M_{GUT}$. However, it is more useful simply to give the
empirical results below :
\begin{eqnarray}
m^2_{e_ie_j}(M_Z) & = & - \sum_{mn} \l{mni}\l{mnj}(M_{GUT})
  (1.7m_0^2 + 0.6M_{1/2}^2 - 0.3 A_0M_{1/2} + 0.6A_0^2 ) \nonumber \\
m^2_{L_iL_j}(M_Z) & = & - \sum_{mn} \l{imn}\l{jmn}(M_{GUT})
  (1.7m_0^2 + 0.6M_{1/2}^2 - 0.3 A_0M_{1/2} + 0.6A_0^2 )  \nonumber \\
     && + \sum_{mn}\lp{imn}\lp{jmn}(M_{GUT})
  (15m_0^2 + 30M_{1/2}^2 -20 A_0M_{1/2} + 5A_0^2 )
\label{lapp}
\end{eqnarray}
The accuracy of these formulae is roughly at the $10\%$ level for the
case where we deal with $\l{}$ only, but is much lower for the case
with $\lp{}$ due to the strong dependence on the QCD coupling, whose
value is uncertain and which typically appears in the RGEs with
larger coefficients. In the latter case we find that the $M_{1/2}^2$
coefficient can vary by almost a factor of two, and so the results
should be regarded as merely indicative of the likely scale of the
terms generated. 

Hence we see that the terms $\Delta
m^2_{\tilde{e}_L\tilde{\mu}_L}/m^2_{\tilde{e}_L}$ and $\Delta
m^2_{\tilde{e}_R\tilde{\mu}_R}/m^2_{\tilde{e}_R}$ are of the  same
order as $\l{1ij}\l{2ij}+3\lp{1ij}\lp{2ij}$ and $\l{ij1}\l{ij2}$
respectively where the couplings are taken at $M_Z$, and are rather
larger when we have non-zero $\lp{}$s than $\l{}$s. Given the range of
values of the various functions and the other couplings, it is then
straightforward to check that we expect the neutralino and chargino
contributions can be at least comparable in magnitude to the direct
contributions, and often larger, particularly for the case of the
$\lp{}$ where the internal particles in the loop for the direct case
are heavier but the impact from $\Delta m^2$ is larger than for the
$\l{}$ case.

We also note that except $H(x)>G(x)$, $L(x)>F(x)$,
and hence if all the mixing matrix elements are significant we expect
the second term in each of the chargino and neutralino contributions to
dominate. Thus we might expect that usually $\tilde A^{\Delta m}$ will
grow with increasing $\tan\beta$, and fall off only slowly with
increasing chargino and neutralino masses, since the explicit mass
factor will partially cancel the dependency in the function.

Another important point which we should make here is that when we add
up all the contributions to $\tilde A_{LR}$ from
equations~(\ref{allr}), (\ref{alplr}), and (\ref{amlr}), and to $\tilde
A_{RL}$ from (\ref{alrl}) and (\ref{amrl}), to obtain the total, there
are likely to be cancellations. For example, if we consider
$\l{i1j}\l{i2j}>0$ in the limit where there is very little
gaugino-higgsino mixing, then the contributions to $\tilde
A^{\lambda}_{LR}$ and the first chargino term of $\tilde A^{\Delta
m}_{LR}$ are both positive, while the contribution to the first
neutralino contribution to $\tilde A^{\Delta m}_{LR}$ is negative, and
other terms will vanish.

\subsection{Numerical Results}
In any case, we can derive through these equations new bounds on
products of $\l{}$ and $\lp{}$. As discussed above for the sneutrino
VEV calculation, we simply impose unification at $M_{GUT}$ with some
pair of R-parity couplings non-zero, evolve down to low energy and then
investigate the amplitudes for $\mu\to e\gamma$. Given that here there
will be sensitivity to virtually every parameter through the
complicated dependence on the spectrum, we shall attempt to pick
typical scenarios and examine whether reasonable conclusions can be
drawn.

Firstly, let us consider the impact of
$\lp{111}(M_{GUT})=\lp{211}(M_{GUT})=0.001$. Here we note that we
expect $\lp{i33}$ to be negligible because of its generically very
large impact on sneutrino VEVs and hence neutrino masses as discussed
above, while we shall avoid $\lp{1i3}$ and $\lp{13i}$ for simplicity.
For $m_0=M_{1/2}=$100GeV, we expect that the values chosen will give a
contribution to the amplitude which is less than the experimental limit
by about two orders of magnitude using Eq.~(\ref{blp}).

We show the resulting contributions to the amplitude as a function of
$M_{1/2}$ in Figure~9. Here we have set $\tan\beta=10$, $m_0=100$GeV,
$A_0=0$, $\mu_4>0$, and $\lp{111}(M_{GUT})=\lp{211}(M_{GUT})=0.001$.
What is remarkable about this figure is that it is clear that in this
case the direct contributions are completely negligible relative to
those from the chargino and neutralino mediated diagrams. Similarly
keeping $M_{1/2}=100$GeV, so that the chargino is always light, and
varying $m_0$, we find contributions of form of Figure~10. Here we find
a curious cancellation between neutralino diagrams, so that the
neutralino contribution at one point changes sign relative to the
chargino contribution, but is in any case always small. Finally we show
the dependence on $A_0$ in Figure~11. Naturally, the direct
contribution to the amplitude is independent of $A_0$ (up to the very
slight dependence of the squark spectrum), but since altering $A_0$ can
affect the chargino and neutralino spectra by altering the values of
$\mu$ and $B$ which give the correct electroweak breaking and also
affect the running of the off-diagonal mass terms, it can have a
significant impact on the indirect contributions.

We now touch upon how sensitive these results are to the other
parameters. Selecting $\mu_4<0$ gives a lighter chargino, which in
general increases the chargino dominated rate, but in practice we find
that since to evade the experimental chargino bound we must then pick
higher values for the soft masses $m_0$ and $M_{1/2}$, this does not in
fact lead to very different behaviour.

Increasing $\tan\beta$ can both directly enhance the contributions from
diagrams with a helicity flip on the chargino or neutralino line and
also change the spectra in a complicated way directly and through
$\mu_4$. In fact, the first of these effects seems most important, as
seen in Figure~12, which is similar to Figure~9, but with $\tan\beta$
set to 30. Here the chargino and neutralino contributions both increase
by around a factor of three, as we would expect if the dominant
contribution to these is via higgsino-neutralino mixed states.

In conclusion of our discussion of the effects generated by a pair of
$\lp{}$ couplings, we have found that the dominant contribution to the
amplitude is that of the chargino followed closely by that of the
neutralino, with the direct contribution from the squark mediated
diagrams almost completely negligible. This clearly has important
implications in any attempt to bound R-parity violating couplings, since
the only case where we can do a reasonably complete analytical analysis
is irrelevant. In general, we can conclude that a choice of
$\lp{1ij}\lp{2ij}(M_{GUT})$ of order $10^{-6}$ is fairly safe for $m_0$
and $M_{1/2}$ of order 100GeV, compared to a limit from the direct
diagrams about two orders of magnitude weaker. However, the various
contributions are sufficently complicated that this can only really be
regarded as an order of magnitude estimate, which may be evaded by
appropriate choice of $\tan\beta$ and the various soft parameters. We
emphasise here that the amplitudes $\tilde A$ are all directly
proportional to the product of the two couplings, and as before it is
straightforward to convert our figures showing the amplitude into
bounds for a given point in parameter space.

Having shown the general behaviour for the $\lp{}$ case, we now
consider two alternative scenarios. Firstly, we consider the case
$\l{131}(M_{GUT})=\l{231}(M_{GUT})=0.01$. Here we note that these
couplings are sufficiently large to give a direct contribution which is
close to the present experimental limit. 

We show the resulting contributions to the amplitude as a function of
$M_{1/2}$ in Figure~13, with $\tan\beta=10$, $m_0=100$GeV, $A_0=0$,
$\mu_4>0$, and $\l{131}(M_{GUT})=\l{231}(M_{GUT})=0.01$. The apparently
pathological behaviour of the direct contribution, which changes sign
even though it does not depend on the stau masses and mixings, can be
understood as follows. From equation~(\ref{allr}), we see that the
amplitude behaves as
\begin{equation}
\tilde{A}_{LR}^{\lambda}  \sim
\frac{1}{m^2_{\tilde\nu_\tau}} - \frac{1}{2m^2_{\tilde{e}_R}}  
\end{equation}
For small $M_{1/2}$, the slepton and sneutrino masses will be similar,
and so the first term on the right hand side of this equation will be
larger and the result positive; however as we increase $M_{1/2}$,
$m^2_{\tilde\nu_\tau}$ will grow larger than $2m^2_{\tilde{e}_R}$ due to
its greater electroweak radiative corrections in the RGEs, at which
point the amplitude changes sign. We find this behaviour occurring when
$m_{\tilde\nu_\tau}$ and $m_{\tilde e_R}$ are around 230GeV and 165GeV
respectively. Here, however, the chargino and neutralino contributions
are still larger than that of the direct contribution, although much
less so than in the case of the $\lp{}$ diagrams.

In Figure~14 we show the same data as for Figure~13 except for
$\tan\beta=2$, but with $M_{1/2}$ held constant at 100GeV, and varying
$A_0$. This is an example of exact cancellation of the amplitude. Since
we often find that the direct and indirect contributions to the
amplitude are of similar magnitude and opposite sign, there are many
cases where the cancellation can be very good, particularly if
$\tan\beta$ is not too large (since if we increase $\tan\beta$ we
increase the chargino and neutralino mediated contributions to the
amplitude without greatly changing the slepton mediated contributions).

Hence, our conclusion for the contribution to $\mu\to e\gamma$ from
setting $\l{131}\l{231}$ non-zero is ultimately that the new
contributions from chargino and neutralino mediated diagrams dominate
for large $\tan\beta$, but that there are some fairly subtle
cancellations which become increasingly significant for smaller
$\tan\beta$. Often the direct and indirect contributions have opposite
sign, and hence it is not really possible to derive a reliable bound.

The last scenario which we shall consider is that of $\tilde A_{RL}$,
which we generate by setting $\l{121}(M_{GUT})=\l{122}(M_{GUT})=0.01$.
Again, we expect the contributions from the direct diagrams to be
relatively close to the experimental limit. However, this case is
rather different from the two previous ones because we no longer have a
chargino contribution. Hence we would expect that the cancellations
between direct and indirect effects might occur for larger $\tan\beta$.
That this is the case is shown in Figure~15, in which we show the
amplitude as a function of $M_{1/2}$ with again $\tan\beta=10$,
$m_0=100$GeV, $A_0=0$, $\mu_4>0$, and
$\l{121}(M_{GUT})=\l{122}(M_{GUT})=0.01$, showing that the cancellation
ensures that the amplitude remains well below the experimental limit.

We conclude this section by summarising our results. Firstly, we find
rather different behaviour for the three different scenarios which we
have considered. For the $\lp{}$ case, the chargino and neutralino
mediated diagrams with flavour violation through soft mass insertions
dominate completely the direct contributions, giving very much tighter
constraints, particularly for large $\tan\beta$. For the $LR$ effects
due to $\l{}$ couplings we find again that the chargino contribution
dominates, but not overwhelmingly, and there can be large
cancellations. For the $RL$ case there are no chargino contributions,
and the neutralino and direct effects are usually of comparable size
and opposite sign. In all of these cases it is very hard to do much
more than gain an idea of the likely size of the effects in question
since the parameter space is complicated, but it is clear that the
$\l{13i}\l{23i}$ and especially $\lp{1ij}\lp{2ij}$ amplitudes will be
much greater than those from the direct contributions alone, while the
neutralino contributions to $\l{ij1}\l{ij2}$ are of the same order of
magnitude as the direct effects. However, since there are so many
possible cancellations between terms, it is essentially impossible to
derive concrete bounds. The strongest reasonable statement is that, for
the values we have considered for pairs of couplings at $M_{GUT}$ of
$\lambda\lambda\simeq 10^{-4}$ and $\lp{}\lp{}\simeq 10^{-6}$ we expect
contributions of order the experimental limit for a very light
spectrum, with the branching ratio scaling as $\lambda^4$ or $\lp{}^4$.

%\section{Quark Flavour Violation}
%Relevant RGEs for QFV with diagonal mass terms dominant :
%\begin{eqnarray}
%16\pi^2 {d m^2_{Q_iQ_j} \over dt}&=& 
%         \sum_{mn} \Bigl(\lp{min}\lp{mjn}
%                  \bigl(m^2_{Q_i} + m^2_{Q_j}
%                     + 2m^2_{d_n} + 2m^2_{L_m}\bigr) \cr
%            && \qquad\qquad + 2\Cp{min}\Cp{mjn}\Bigr) \\
%16\pi^2 {d m^2_{u_iu_j} \over dt}&=& 
%         \sum_{mn} \Bigl(\lpp{imn}\lpp{jmn}
%                 \bigl(m^2_{u_i} + m^2_{u_j} + 4m^2_{d_n}\bigr) \cr
%            && \qquad\qquad + 2\Cpp{imn}\Cpp{jmn}\Bigr) \\
%16\pi^2 {d m^2_{d_id_j} \over dt}&=& 
%         \sum_{mn}\Bigl(2\lp{mni}\lp{mnj}
%                 \bigl(m^2_{d_i} + m^2_{d_j}
%                    + 2m^2_{Q_n} + 2m^2_{L_m}\bigr) \cr
%            && \qquad\qquad + 4\Cpp{mni}\Cpp{mnj} \Bigr)\cr
%      &+&\sum_{mn}\Bigl(2\lpp{mni}\lpp{mnj}
%                 \bigl(m^2_{d_i}+m^2_{d_j}
%                    + 2m^2_{d_n} + 2m^2_{u_m}\bigr) \cr
%            && \qquad\qquad + 4\Cpp{mni}\Cpp{mnj}\Bigr)
%\end{eqnarray}

\section{Conclusion}
In this paper we have given full RGEs for the MSSM with R-parity
violation with the inclusion of all soft terms as well as all
dimensionless couplings. We have given solutions to these equations for
small R-parity violating couplings to relate values at the unification
scale to those at low energy, and given the triviality bounds on
couplings.

The inclusion of R-parity violation in our superpotential through
dimensionless terms allows the generation of lepton-Higgs mixing which
leads to sneutrino VEVs and hence neutrino masses. We have presented a
detailed discussion of sneutrino VEV generation, together with a
calculation assuming universal soft terms respecting lepton number at
the unification scale. This shows that the indirect generation of
sneutrino VEVs through the running of the RGEs for the soft terms often
leads to larger effects than those derived directly from one loop
diagrams, and hence that $\l{133}$, $\l{233}$, $\lp{i33}$ must all be
small. Typically we find that values of $\l{i33}$ and $\lp{i33}$ of
order $10^{-2}$ and $10^{-3}$ at the GUT scale give masses to the
corresponding neutrino of order hundreds to thousands of eV, although
the exact value is quite dependent on the unification scale parameters.

Similarly, R-parity violation at the unification scale can generate
large lepton and quark flavour violation both directly through one loop
diagrams and indirectly through the generation of off-diagonal soft
masses and hence flavour violation in the scalar sector. We have
studied the process $\mu\to e\gamma$, which we have shown to be very
strongly affected by chargino and neutralino mediated diagrams. These
typically dominate the direct contributions which had already been
calculated, typically by several orders of magnitude for the case of
the $\lp{}$ couplings, but there are often cancellations so that it is
not possible to give precise bounds on the couplings from such
processes. However, unless we invoke arbitrary cancellations, the
typical size of such indirect effects on FCNC are likely to be the
dominant constraint on the building of a model with non-zero R--parity
violating couplings.

Our main conclusion from both these calculations is that R-parity
violation can generate large lepton flavour and lepton number violating
effects through the running of the dimensionful RGEs, and that these
effects are often much larger than those which are generated directly
by the couplings themselves. However, it is virtually impossible to
turn this statement into hard numerical bounds because of the number of
free input parameters and the large cancellations between different
contributions.

\noindent
{\bf Acknowledgements \hfil} \\
We would like to thank Roger Phillips and Mike Berger for discussing
their work of reference \cite{roger}. BdC thanks J.R. Espinosa for
very enlightening discussions. PLW is very grateful to the University 
of Sussex for their hospitality during the final stages of this work. 
The work of BdC was supported by a PPARC Postdoctoral Fellowship.

\appendix
\section{Renormalisation Group Equations}
We now present our conventions and RG equations. The superpotential
and soft potential are given by
\begin{eqnarray}
W&=&
   h_tQ_3H_2u_3 + h_b Q_3H_1d_3 + h_\tau L_3H_1e_3 \cr 
   &&+ \frac{1}{2}\l{ijk}L_iL_je_k + \lp{ijk}L_iQ_jd_k
                + \frac{1}{2}\lpp{ijk}u_id_jd_k \cr
   &&+ \mu_4 H_1H_2 + \mu_iL_iH_2 \cr
V_{\rm soft}&=&
   \eta_tQ_3H_2u_3 + \eta_b Q_3H_1d_3 + \eta_\tau L_3H_1e_3 + h.c. \cr
   &&+ \frac{1}{2}\C{ijk}L_iL_je_k + \Cp{ijk}L_iQ_jd_k
                + \frac{1}{2}\Cpp{ijk}u_id_jd_k + h.c.  \cr
   &&+ \frac{1}{2}M_a\lambda_a^c\lambda_a
                + \sum_{a,b} m_{ab}^2 \varphi_a\bar\varphi_b \cr
   &&+ D_4 H_1H_2 + D_iL_iH_2 + h.c.
\end{eqnarray}
Here the notation is conventional, with the exception that we do not
write the trilinear terms as $A\times$coupling since this is not
convenient algebraically in the context of R--parity violation, and
similarly we do not write the bilinear soft term as $B\mu$. It should
be noted that we now label $\mu$ as $\mu_4$ so as to make clear the
analogy between the R--parity violating bilinear terms $\mu_i$ and
$\mu_4$. Indices indicate generation, and all gauge indices are
suppressed. Raising and lowering generation indices has no particular
significance. In our expressions we will use $\mu$ interchangably with
$\mu_4$ and $A_\mu$ for $\eta_\mu/h_\mu$, and we express the mass
eigenstates as $\tilde e^{(1)}$, $\tilde e^{(2)}$, and similarly for
quarks.

The derivation of the RG equations is straightforward using standard
techniques found for example in \cite{ds,mv}. Here, we use the
convention that all indices which do not appear on the left hand side
of the equations are summed over for the three generations, and we
define
\begin{equation}
\xi=\sum_{a}Y_a g_1^2 m_{aa}^2
\end{equation}
where the sum is over all particles, and must include all appropriate
degrees of freedom factors from, for example, colour and generation.
Weak hypercharge is here normalised so that $Y_Q=1/6$. We use
$t=\ln\mu$, where $\mu$ is the $\overline{MS}$ renormalisation scale.
\begin{eqnarray}
16\pi^2 {d h_t \over dt}&=& 
   h_t \bigl( 6h_t^2 +h_b^2 + \lp{m3n}^2 + \lpp{3mn}^2 \bigr )\cr
       &&  - h_t \bigl ({13\over 9}g_1^2 + 3g_2^2 + {16\over 3}g_3^2\bigr )
    \cr
16\pi^2 {d h_b \over dt}&=& 
   h_b \bigl( 6h_b^2 + h_t^2 + h_\tau^2
       + 3\lp{m33}^2 + \lp{m3n}^2 + 2\lp{mn3}^2 + 2\lpp{mn3}^2 \bigr ) \cr
       && + h_\tau \l{m33}\lp{m33}
         - h_b \bigl( {7\over 9}g_1^2 + 3g_2^2 + {16\over 3}g_3^2\bigr )
    \cr
16\pi^2 {d h_\tau \over dt}&=& 
   h_\tau \bigl( 4h_\tau^2 + 3h_b^2 
       +\l{m33}^2 + \l{m3n}^2 +\l{mn3}^2 + 3\lp{3mn}^2 \bigr ) \cr
       && + 3h_b \l{m33}\lp{m33}
         - h_\tau \bigl( 3g_1^2 + 3g_2^2\bigr )
    \\
16\pi^2 {d \l{ijk} \over dt}&=& 
      \l{lmn} \bigl( \l{imn}\l{ljk} + \l{imk}\l{ljn} +\l{ijn}\l{lmk} \bigr ) \cr
       &&  +\lp{lmn} \bigl( 3\l{ljk}\lp{imn} + 3\l{ilk}\lp{jmn} \bigr ) \cr
       &&  +\l{ijk} h_\tau^2 \bigl( \delta_{i3} + \delta_{j3} + 2\delta_{k3}
                + \delta_{j3}\delta_{k3} + \delta_{i3}\delta_{k3} \bigr ) \cr
       &&  +h_b h_\tau \bigl( 3\delta_{j3}\delta_{k3}\lp{i33}
                 - 3\delta_{i3}\delta_{k3}\lp{j33} \bigr ) \cr
       &&  -\l{ijk} \bigl( 3g_1^2 + 3g_2^2\bigr )
    \cr
16\pi^2 {d \lp{ijk} \over dt}&=& 
      \lp{lmn} \bigl( 3\lp{imn}\lp{ljk} + 2\lp{ijn}\lp{lmk} 
                          + \lp{imk}\lp{ljn} \bigr ) \cr
       &&  + \l{lmn}\lp{ljk}\l{imn} + 2\lpp{lmn}\lp{ijn}\lpp{lmk} \cr
       &&  +\lp{ijk} \bigl( \delta_{i3}h_\tau^2 
                          + \delta_{j3}h_t^2 + \delta_{j3}h_b^2 
                          + 2\delta_{k3}h_b^2
                          + 3\delta_{j3}\delta_{k3}h_b^2 \bigr )
                          + \delta_{j3}\delta_{k3}h_b h_\tau\l{i33} \cr
       &&  -\lp{ijk} \bigl( {7\over 9}g_1^2 + 3g_2^2 + {16\over 3}g_3^2\bigr )
    \cr
16\pi^2 {d \lpp{ijk} \over dt}&=& 
      \lpp{lmn} \bigl( \lpp{imn}\lpp{ljk} + 2\lpp{imk}\lpp{ljn} 
                          + 2\lpp{ijn}\lpp{lmk} \bigr ) \cr
       &&  +\lp{lmn} \bigl( 2\lpp{ink}\lp{lmj} + 2\lpp{ijn}\lp{lmk} \bigr ) \cr
       &&  +\lpp{ijk} \bigl( 2\delta_{i3}h_t^2 + 2\delta_{j3}h_b^2 
                          + 2\delta_{k3}h_b^2 \bigr ) \cr
       &&  -\lpp{ijk} \bigl( {4\over 3}g_1^2 + 8g_3^2 \bigr )
    \\
16\pi^2 {d \eta_t \over dt}&=& 
   \eta_t \bigl( 18h_t^2 +h_b^2 + \lp{m3n}^2 + \lpp{3mn}^2 \bigr )\cr
       &&  + h_t \bigl ( 2h_b\eta_b + 2\lp{m3n}\Cp{m3n}
                        + 2\lpp{3mn}\Cpp{3mn} \bigr ) \cr
       &&  - \eta_t \bigl ({13\over 9}g_1^2 + 3g_2^2 + {16\over 3}g_3^2\bigr )
          + 2h_t \bigl ({13\over 9}g_1^2M_1 + 3g_2^2M_2
                       + {16\over 3}g_3^2M_3 \bigr )
      \cr
16\pi^2 {d \eta_b \over dt}&=& 
   \eta_b \bigl( 18h_b^2 + h_\tau^2 + h_t^2 + 6\lp{m33}^2
            + 2\lp{mn3}^2 + \lp{m3n}^2 +2\lpp{mn3}^2 \bigr ) \cr
       && + \eta_\tau \bigl (2h_bh_\tau + 2\l{m33}\lp{m33} \bigr )
              + 2h_bh_t\eta_t
              + \Cp{m33} \bigl (3h_b\lp{m33}+h_\tau\l{m33} \bigr ) \cr
       && + 4 h_b\lp{mn3}\Cp{mn3} + 2 h_b\lp{m3n}\Cp{m3n}
             + 4 h_b\lpp{mn3}\Cpp{mn3} \cr
       &&  -\eta_b \bigl( {7\over 9}g_1^2 + 3g_2^2 + {16\over 3}g_3^2\bigr )
          + 2h_b \bigl( {7\over 9}g_1^2M_1 + 3g_2^2M_2 
                         + {16\over 3}g_3^2M_3 \bigr )
      \cr
16\pi^2 {d \eta_\tau \over dt}&=& 
   \eta_\tau \bigl( 12h_\tau^2 + 3h_b^2
            + 2\l{m33}^2 + \l{m3n}^2 +\l{mn3}^2 + 3\lp{3mn}^2 \bigr ) \cr
       && + \eta_b \bigl (6h_\tau h_b + 6\l{m33}\lp{m33} \bigr )
         + \C{m33} \bigl (3h_b\lp{m33}+h_\tau\l{m33} \bigr ) \cr
       && + 2 h_\tau\l{m3n}\C{m3n} + 2 h_\tau\l{mn3}\C{mn3}
         + 6 h_\tau\lp{3mn}\Cp{3mn} \bigr )  \cr
       &&  -\eta_\tau \bigl( 3g_1^2 + 3g_2^2\bigr )
          +2h_\tau \bigl( 3g_1^2M_1 + 3g_2^2M_2 \bigr )
      \\
16\pi^2 {d \C{ijk} \over dt}&=& 
      \l{lmn} \bigl( \l{imn}\C{ljk} + \l{ljn}\C{imk} +\l{lmk}\C{ijn} \cr
       && \qquad     + 2\l{ijn}\C{lmk} + 2\l{ljk}\C{imn}
                         + 2\l{imk}\C{ljn} \bigr ) \cr
       &&  +\lp{lmn} \bigl( 3\lp{imn}\C{ljk} + 3\lp{jmn}\C{ilk} 
                         + 6\l{ljk}\Cp{imn} + 6\l{ilk}\Cp{jmn} \bigr ) \cr
       &&  +\C{ijk} h_\tau^2 \bigl( \delta_{i3} + \delta_{j3} + 2\delta_{k3}
                + 2\delta_{j3}\delta_{k3} + 2\delta_{i3}\delta_{k3} \bigr ) \cr
       &&  +\l{ijk} h_\tau \eta_\tau \bigl( 2\delta_{i3}
                + 2\delta_{j3} + 4\delta_{k3}
                + \delta_{j3}\delta_{k3} + \delta_{i3}\delta_{k3} \bigr ) \cr
       &&  + h_b \eta_\tau \bigl( 3\delta_{j3}\delta_{k3}\lp{i33}
                 - 3\delta_{i3}\delta_{k3}\lp{j33} \bigr ) \cr
       &&  + 2h_b h_\tau \bigl( 3\delta_{j3}\delta_{k3}\Cp{i33}
                 - 3\delta_{i3}\delta_{k3}\Cp{j33} \bigr ) \cr
       &&  -\C{ijk} \bigl( 3g_1^2 + 3g_2^2\bigr )
          +2\l{ijk} \bigl( 3g_1^2M_1 + 3g_2^2M_2 \bigr )
    \cr
16\pi^2 {d \Cp{ijk} \over dt}&=& 
      \lp{lmn} \bigl( 3\lp{imn}\Cp{ljk} + \lp{ljn}\Cp{imk}
                  + 2\lp{lmk}\Cp{ijn} \cr
       && \qquad   + 4\lp{ijn}\Cp{lmk} + 2\lp{imk}\Cp{ljn}
                  + 6\lp{ljk}\Cp{imn} \bigr ) \cr
       &&  +\l{lmn} \bigl( \l{imn}\Cp{ljk} + 2\lp{ljk}\C{imn} \bigr ) \cr
       &&  +\lpp{lmn} \bigl( 2\lpp{lmk}\Cp{ijn} + 4\lp{ijn}\Cpp{lmk} \bigr ) \cr
       &&  +\Cp{ijk} \bigl( \delta_{i3}h_\tau^2 + \delta_{j3}h_t^2
                  + \delta_{j3}h_b^2 + 2\delta_{k3}h_b^2 
                  + 6\delta_{j3}\delta_{k3}h_b^2\bigr ) \cr
       &&  +\lp{ijk} \bigl( 2\delta_{i3}h_\tau\eta_\tau + 2\delta_{j3}h_t\eta_t
                  + 2\delta_{j3}h_b\eta_b + 4\delta_{k3}h_b\eta_b
                  + 3\delta_{j3}\delta_{k3}h_b\eta_b \bigr ) \cr
       &&  + \bigl( \delta_{j3}\delta_{k3}h_\tau\l{i33}\eta_b
                 + 2\delta_{j3}\delta_{k3}h_b h_\tau\C{i33} \bigr ) \cr
       &&  - \Cp{ijk} \bigl( {7\over 9}g_1^2 + 3g_2^2 + {16\over 3}g_3^2\bigr )
                 + 2\lp{ijk} \bigl( {7\over 9}g_1^2M_1 + 3g_2^2M_2 
                 + {16\over 3}g_3^2M_3 \bigr )
    \cr
16\pi^2 {d \Cpp{ijk} \over dt}&=& 
      \lpp{lmn} \bigl( \lpp{imn}\Cpp{ljk} + 2\lpp{ljn}\Cpp{imk}
                                     + 2\lpp{lmk}\Cpp{ijn} \cr
       && \qquad     + 4\lpp{ijn}\Cpp{lmk} + 4\lpp{imk}\Cpp{ljn}
                                    + 2\lpp{ljk}\Cpp{imn} \bigr ) \cr
       &&  +\lp{lmn} \bigl( 2\lp{lmj}\Cpp{ink} + 2\lp{lmk}\Cpp{ijn} 
                         + 4\lpp{ijn}\Cp{lmk} + 4\lpp{ink}\Cp{lmj} \bigr ) \cr
       &&  +\Cpp{ijk} \bigl( 2\delta_{i3}h_t^2 + 2\delta_{j3}h_b^2
                  + 2\delta_{k3}h_b^2 \bigr ) \cr
       &&  +\lpp{ijk} \bigl( 4\delta_{i3}h_t\eta_t + 4\delta_{j3}h_b\eta_b
                 + 4\delta_{k3}h_b\eta_b \bigr ) \cr
       &&  - \Cpp{ijk} \bigl( {4\over 3}g_1^2 + 8g_3^2\bigr )
                 + \lpp{ijk} \bigl( {8\over 3}g_1^2M_1 + 16g_3^2M_3 \bigr )
    \\
16\pi^2 {d m^2_{e_ie_j} \over dt}&=& 
           \l{mni}\l{mnk}m^2_{e_je_k} + \l{mnj}\l{mnk}m^2_{e_ie_k}
                                      + 4\l{lmi}\l{lnj}m^2_{L_mL_n} \cr
      && + m^2_{e_ie_j} \bigl ( 2\delta_{i3}h_\tau^2
                                        + 2\delta_{j3}h_\tau^2 \bigr )
        + \delta_{i3}\delta_{j3} \bigl ( 4h_\tau^2m_{H_1}^2
                        + 4h_\tau^2m_{L_3L_3}^2 + 4\eta_\tau^2 \bigr ) \cr
      &&  + 2\C{mni}\C{mnj} + 2\delta_{ij}\xi - 8\delta_{ij}g_1^2M_1^2
     \cr
16\pi^2 {d m^2_{L_iL_j} \over dt}&=& 
           \l{imn}\l{kmn}m^2_{L_jL_k} + \l{jmn}\l{kmn}m^2_{L_iL_k} \cr
      && + 3\lp{imn}\lp{kmn}m^2_{L_jL_k} + 3\lp{jmn}\lp{kmn}m^2_{L_iL_k} \cr
      && + 2\l{ilm}\l{jln}m^2_{e_me_n} + 2\l{iml}\l{jnl}m^2_{L_mL_n} \cr
      && + 6\lp{ilm}\lp{jln}m^2_{d_md_n} + 6\lp{iml}\lp{jnl}m^2_{Q_mQ_n} \cr
      && + m^2_{L_iL_j} \bigl ( \delta_{i3}h_\tau^2
         + \delta_{j3}h_\tau^2 \bigr )
         + \delta_{i3}\delta_{j3} \bigl ( 2h_\tau^2m_{H_1}^2
                     + 2h_\tau^2m_{e_3e_3}^2 + 2\eta_\tau^2 \bigr ) \cr
      && + 2\C{imn}\C{jmn} + 6\Cp{imn}\Cp{jmn}
        - \delta_{ij}\xi - 2\delta_{ij}\bigl(g_1^2M_1^2+3g_2^2M_2^2\bigr )
     \cr
16\pi^2 {d m^2_{Q_iQ_j} \over dt}&=& 
           \lp{min}\lp{mkn}m^2_{Q_jQ_k} + \lp{mjn}\lp{mkn}m^2_{Q_iQ_k} \cr
      && + 2\lp{kim}\lp{kjn}m^2_{d_md_n} + 2\lp{mik}\lp{njk}m^2_{L_mL_n}
            + 2\Cp{min}\Cp{mjn} \cr
      && + m^2_{Q_iQ_j} \bigl ( \delta_{i3}h_t^2 + \delta_{i3}h_b^2
                  + \delta_{j3}h_t^2 + \delta_{j3}h_b^2 \bigr ) \cr
      && + \delta_{i3}\delta_{j3} \bigl ( 
                    2h_t^2m_{H_2}^2 + 2h_t^2m_{u_3u_3}^2 + 2\eta_t^2
                  + 2h_b^2m_{H_1}^2 + 2h_b^2m_{d_3d_3}^2 + 2\eta_b^2 \bigr ) \cr
      && + {1\over 3}\delta_{ij}\xi
        - 2\delta_{ij}\bigl({1\over 9}g_1^2M_1^2 + 3g_2^2M_2^2
                  + {16\over 3}g_3^2M_3^2 \bigr )
     \cr
16\pi^2 {d m^2_{u_iu_j} \over dt}&=& 
           \lpp{imn}\lpp{kmn}m^2_{u_ju_k} + \lpp{jmn}\lpp{kmn}m^2_{u_iu_k}
                    + 4\lpp{ikm}\lpp{jkn}m^2_{d_md_n} + 2\Cpp{imn}\Cpp{jmn} \cr
      && + m^2_{u_iu_j} \bigl ( 2\delta_{i3}h_t^2 + 2\delta_{j3}h_t^2 \bigr )
         + \delta_{i3}\delta_{j3} \bigl ( 
                    4h_t^2m_{H_2}^2 + 4h_t^2m_{Q_3Q_3}^2 + 4\eta_t^2 \bigr ) \cr
      && - {4\over 3}\delta_{ij}\xi
         - 2\delta_{ij}\bigl({16\over 9}g_1^2M_1^2
         + {16\over 3}g_3^2M_3^2 \bigr )
     \cr
16\pi^2 {d m^2_{d_id_j} \over dt}&=& 
         2\lp{mni}\lp{mnk}m^2_{d_jd_k} + 2\lp{mnj}\lp{mnk}m^2_{d_id_k} \cr
      && + 2\lpp{mni}\lpp{mnk}m^2_{d_jd_k} + 2\lpp{mnj}\lpp{mnk}m^2_{d_id_k} \cr
      && + 4\lp{kmi}\lp{knj}m^2_{Q_mQ_n}
                    + 4\lp{mki}\lp{nkj}m^2_{L_mL_n} \cr
      && + 4\lpp{kmi}\lpp{knj}m^2_{d_md_n}
                    + 4\lpp{mki}\lpp{nkj}m^2_{u_mu_n} \cr
      && + 4\Cp{mni}\Cp{mnj} + 4\Cpp{mni}\Cpp{mnj} \cr
      && + m^2_{d_id_j} \bigl ( 2\delta_{i3}h_b^2 + 2\delta_{j3}h_b^2 \bigr )
        + \delta_{i3}\delta_{j3} \bigl ( 
                    4h_b^2m_{H_1}^2 + 4h_b^2m_{Q_3Q_3}^2 + 4\eta_b^2 \bigr ) \cr
      && + {2\over 3}\delta_{ij}\xi
        - 2\delta_{ij}\bigl({4\over 9}g_1^2M_1^2 + {16\over 3}g_3^2M_3^2 \bigr )
\end{eqnarray}

In the above equations, we have not allowed for the inclusion of Yukawa
terms other than those of the third generation, and we have also not
presented dimensionful terms such as $m_{L_iH_1}^2$ which can be
generated by the mixing of $L_i$ and $H_1$. However, these may be
trivially derived from the above by simply regarding $H_1$ as $L_4$,
dropping the explicit $h_b$ and $h_\tau$ dependence of the equations,
and setting
\begin{equation}
\l{343}=-\l{433}=h_{\tau}
\qquad
\lp{433}=-h_b
\qquad
\C{343}=-\C{433}=\eta_{\tau}
\qquad
\Cp{433}=-\eta_b
\end{equation}
Adopting this notation, we find
\begin{eqnarray}
16\pi^2 \frac{d\mu_i}{dt}&=& 
      \mu_i(3h_t^2 - g_1^2 - 3g_2^2)
        + 3\lp{ikl}\lp{jkl}\mu_j 
        + \l{ikl}\l{jkl}\mu_j  \nonumber \\
16\pi^2 \frac{dD_i}{dt}&=& 
      D_i(3h_t^2 - g_1^2 - 3g_2^2)
         + 3\lp{ikl}\lp{jkl}D_j 
         + \l{ikl}\l{jkl}D_j  \\
       && + 6h_t\eta_t\mu_i + 6 \Cp{ikl}\lp{jkl}\mu_j
         + 2\C{ikl}\l{jkl}\mu_j
         + 2g_1^2M_1\mu_i + 6g_2^2M_2\mu_i 
\nonumber
\end{eqnarray}
We further note that although we have given the RGEs for the third
generation Yukawas only, it is straightforward to use this notation to
generate the full RGEs including general mass matrices and the full CKM
dependence through
\begin{equation}
\lp{4ij}=-h^d_{ij}
\end{equation}
where $h^d_{ij}$ is the full Yukawa matrix for the down-type quarks,
which can at low energies be found from the quark masses and the CKM
matrix as defined, for example, in \cite{pdb}. In general, the R-parity
violating couplings will prevent all the mass matrices remaining
diagonal even if they are initially so chosen, as discussed above.

For completeness, we also give the RGE for $m^2_{H_2H_2}$, which is
unchanged from that of the MSSM.
\begin{eqnarray}
16\pi^2 {d m^2_{H_2H_2} \over dt}&=& 
         6h_t^2\bigl( m_{Q_3Q_3}^2 + m_{H_2H_2}^2 + m_{u_3u_3}^2 \bigr )
             + 6\eta_t^2 \qquad \cr
           && - \bigl( 2g_1^2M_1^2 + 6g_2^2M_2^2 \bigr )
             + \xi 
\end{eqnarray}

\section{Definition of Functions}
The expressions for the amplitudes in $\mu\to e\gamma$ presented
earlier employ a number of functions, most of which can be found in
either ref.~\cite{us} or in Appendix B of ref.~\cite{bert}. We
repeat the definitions here for convenience.
\begin{eqnarray}
F_1(x)&=&\frac{1}{12(x-1)^4} [ x^3-6x^2+3x+2+6x\ln(x)] \cr
F_2(x)&=&\frac{1}{x}F_1(\frac{1}{x}) \cr
      &=&\frac{1}{12(x-1)^4} [ 2x^3+3x^2-6x+1-6x^2\ln(x)] \cr
F_3(x)&=&\frac{1}{2(x-1)^3} [ x^2-4x+3+2\ln(x)] \cr
F_4(x)&=&\frac{1}{2(x-1)^3} [ x^2-1-2x\ln(x)] \cr
G(x) &=& F_1(x)+xF_1^{\prime}(x)\cr
     &=&\frac{1}{6(x-1)^5} [ x^3+9x^2-9x-1-6x(x+1)\ln(x)] \cr
F(x) &=& F_2(x)+xF_2^{\prime}(x)\cr
     &=&\frac{1}{12(x-1)^5} [ -17x^3+9x^2+9x-1+6x^2(x+3)\ln(x)] \cr
H(x) &=& F_3(x)+xF_3^{\prime}(x)\cr
     &=&\frac{1}{2(x-1)^4} [ x^2+4x-5-2(2x+1)\ln(x)] \cr
L(x) &=& F_4(x)+xF_4^{\prime}(x)\cr
     &=&\frac{1}{2(x-1)^4} [ -5x^2+4x+1+2x(x+2)\ln(x)]
\end{eqnarray}
Here the prime simply indicates the first derivative. Certain asymptotic
values of these functions are useful, as listed in Table~2 below.

\begin{center}
\begin{tabular}{|c|c|c|c|c|c|c|c|c|}\hline
$x$ & $F_1(x)$ & $F_2(x)$ & $F_3(x)$ & $F_4(x)$ &
 $G(x)$ & $F(x)$ & $H(x)$ & $L(x)$ \\ \hline
0 & $1/6$ & $1/12$ & $-\ln(x)$ & $1/2$ & 
 $1/6$ & $1/12$ & $-\ln(x)-5/2$ & $1/2$
 \\ \hline
1 & $1/24$ & $1/24$ & $1/3$ & $1/6$ & 
 $1/60$ & $1/40$ & $1/12$ & $1/12$
 \\ \hline
$\infty$ & $1/12x$ & $1/6x$ & $1/2x$ & $1/2x$ & 
 $1/6x^2$ & $\ln(x)/2x^2$ & $1/2x^2$ & $\ln(x)/x^2$
 \\ \hline
\end {tabular}
\vskip .5cm
\footnotesize{
Table 2. Values of the various functions in useful limits.}
\end{center}

\newpage

\section{Figure Captions}

\noindent
{\bf Figure 1:}
Feynman diagrams contributing to $L_i-H_1$ mixing and hence to the
generation of sneutrino VEVs. We do not show possible mass insertions
on the lines.

\noindent
{\bf Figure 2a:}
Absolute values of the sneutrino VEV (dashed lines) and corresponding
neutrino mass (solid lines) as a function of $m_{L_i}$, with parameters
$m_t=175$GeV, $\alpha_3(M_Z)=0.12$, $\tan\beta=2$ and 20,
$M_{1/2}=500$GeV, $A_0=0$, both signs of $\mu$, and
$\l{133}(M_{GUT})=0.01$.

\noindent
{\bf Figure 2b:}
Absolute values of the sneutrino VEV (dashed lines) and corresponding
neutrino mass (solid lines) as a function of $m_{L_i}$, with parameters
$m_t=175$GeV, $\alpha_3(M_Z)=0.12$, $\tan\beta=$2 and 20,
$M_{1/2}=500$GeV, $A_0=0$, both signs of $\mu$, and
$\lp{133}(M_{GUT})=0.001$.

\noindent
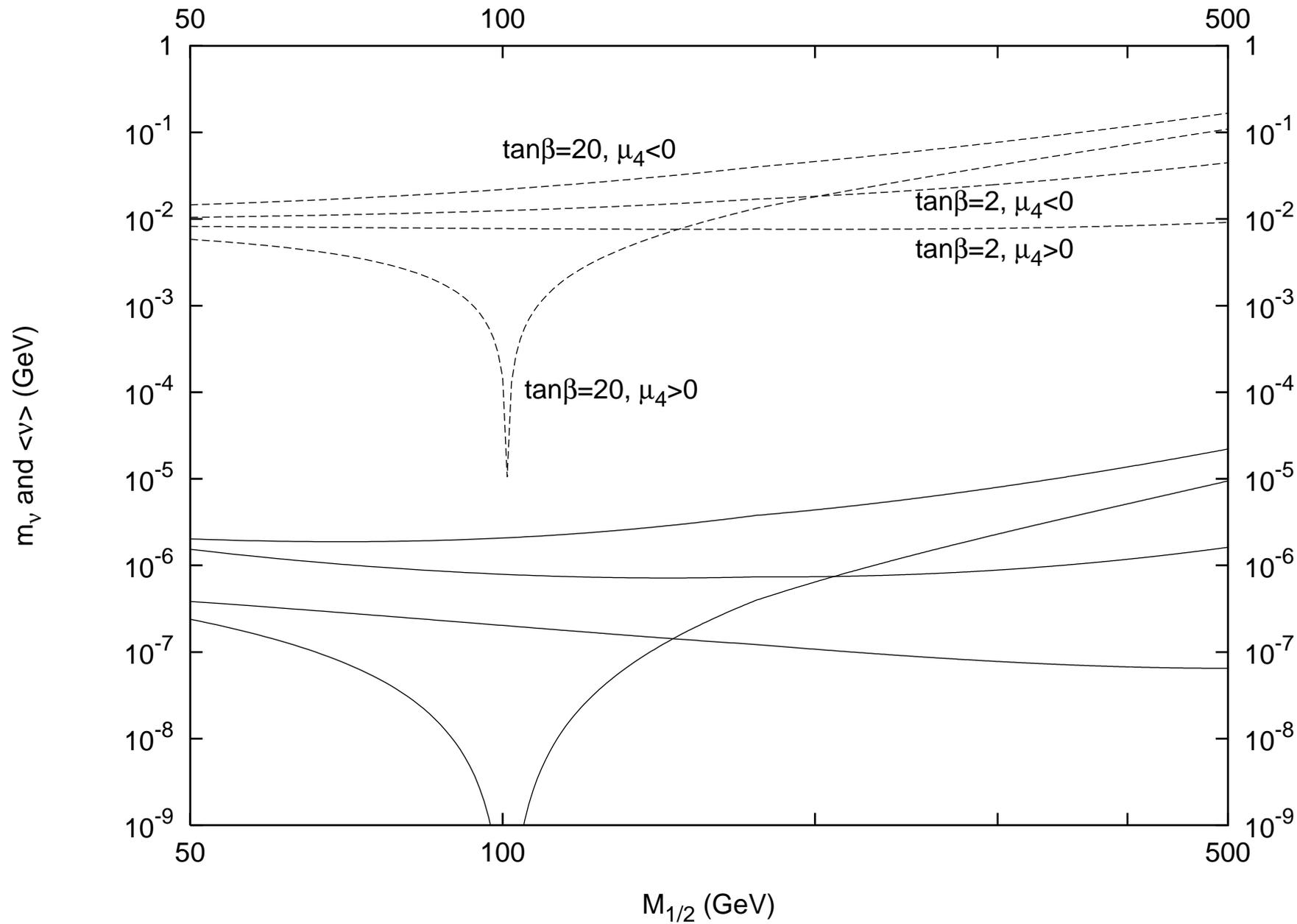
{\bf Figure 3:}
Absolute values of the sneutrino VEV (dashed lines) and corresponding
neutrino mass (solid lines) as a function of $M_{1/2}$, with parameters
$m_t=175$GeV, $\alpha_3(M_Z)=0.12$, $\tan\beta=2,\ 20$,
$m_{L_i}=500$GeV, $A_0=0$, both signs of $\mu$, and
$\lp{133}(M_{GUT})=0.001$.

\noindent
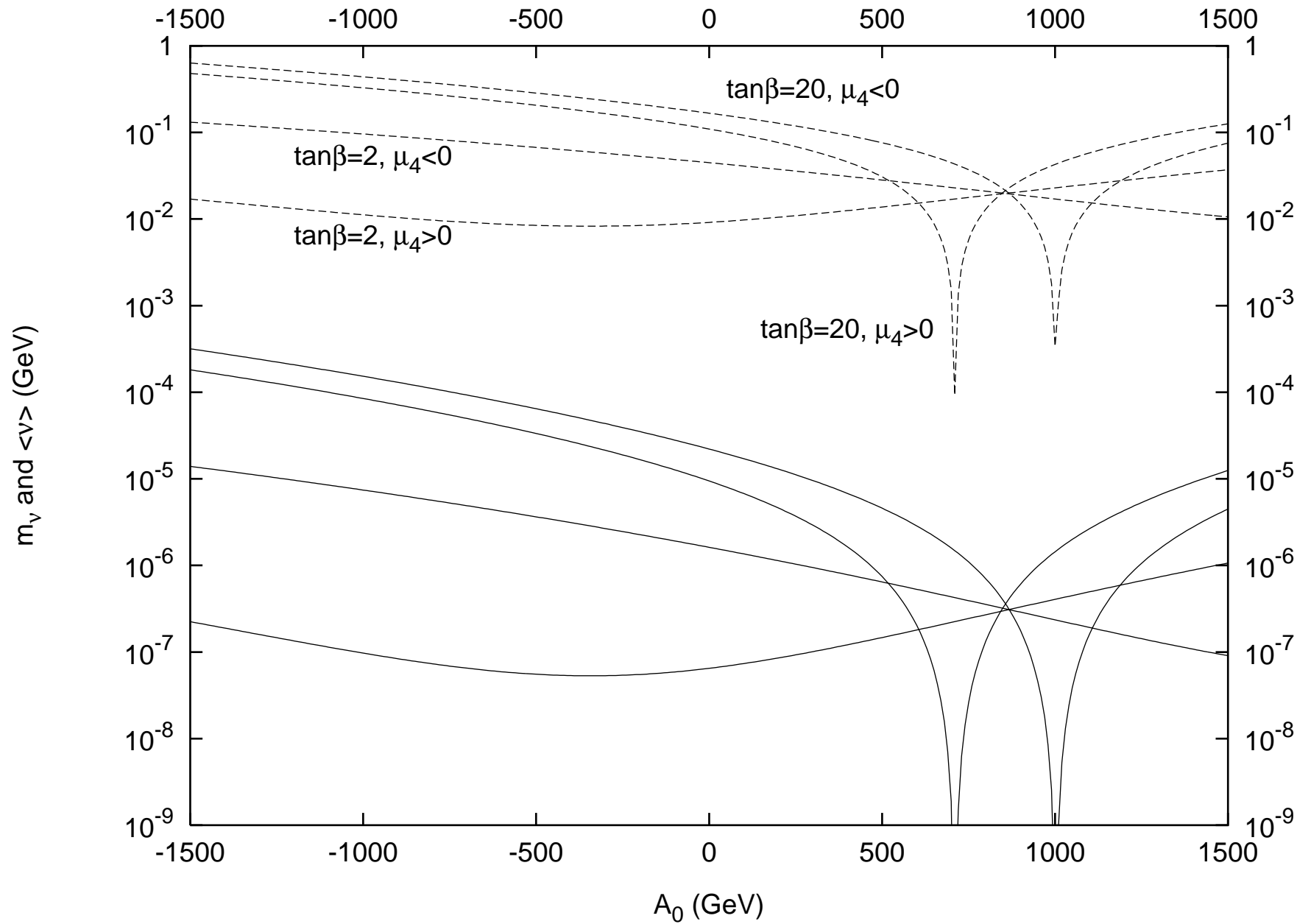
{\bf Figure 4:}
Absolute values of the sneutrino VEV (dashed lines) and corresponding
neutrino mass (solid lines) as a function of $A_0$, with parameters
$m_t=175$GeV, $\alpha_3(M_Z)=0.12$, $\tan\beta=2$ and 20,
$M_{1/2}=m_{L_i}=500$GeV, both signs of $\mu$, and
$\lp{133}(M_{GUT})=0.001$.

\noindent
{\bf Figures 5:}
Diagrams contributing to $\tilde A^{\lambda}_{LR}$. The analagous
contributions to $\tilde A^{\lambda}_{RL}$ have a left handed internal
lepton line. All diagrams are presented without photon lines for
clarity, and the crosses indicate helicity flips except where otherwise
marked.

\noindent
{\bf Figure 6:}
Diagrams contributing to $\tilde A^{\lambda^{\prime}}_{LR}$. There are
no analagous diagrams for $\tilde A^{\lambda^{\prime}}_{RL}$.

\noindent {\bf Figure 7:}
Diagrams contributing to $\tilde A^{\Delta m}_{LR}$. The analagous
contribution to $\tilde A^{\Delta m}_{RL}$ is proportional to the
electron mass in the chargino case and hence neglected, while for the
$RL$ neutralino case right and left handed fields are interchanged.

\noindent
{\bf Figures 8:}
Examples of diagrams which contribute to $\tilde A^{\lambda}_{LR}$ with
the insertion of a sneutrino VEV and hence neutralino-neutrino or
slepton-chargino mixing.

\noindent
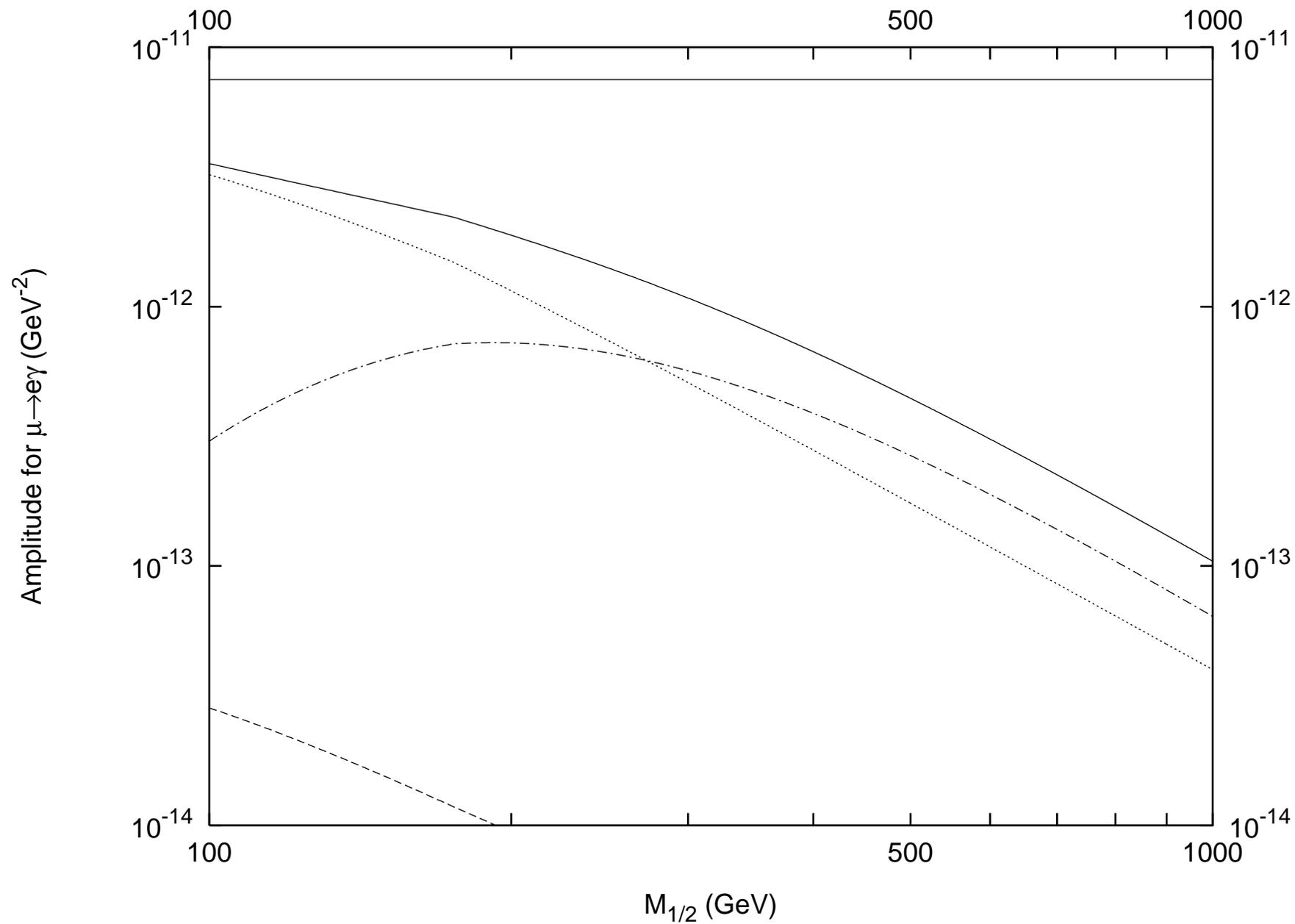
{\bf Figure 9:}
Absolute values of amplitudes for $\mu\to e\gamma$ from direct R-parity
violation diagrams (dashed lines), neutralino mediated diagrams
(dot-dashed lines), and chargino mediated diagrams (dotted lines)
plotted against $M_{1/2}$. We also show the total amplitude (solid
line) and the experimental bound on the amplitude (horizontal solid
line). Parameters are $m_t=175$GeV, $\alpha_3(M_Z)=0.12$,
$\tan\beta=10$, $m_0=100$GeV, $A_0=0$, $\mu_4>0$, and
$\lp{111}(M_{GUT})=\lp{211}(M_{GUT})=0.001$.

\noindent
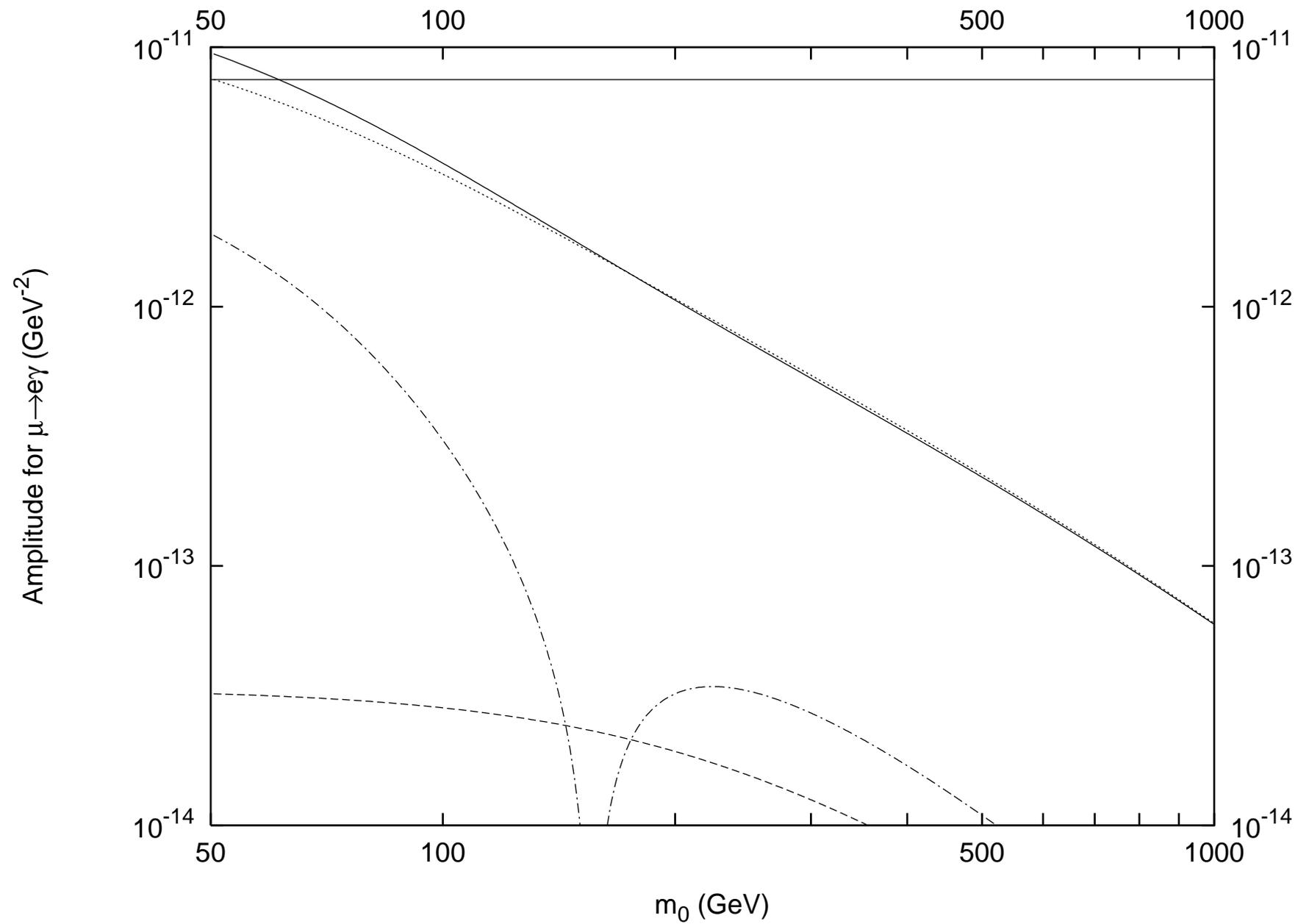
{\bf Figure 10:}
Exactly as for Figure 9, but varying $m_0$ and with $M_{1/2}$ set to
100GeV.

\noindent
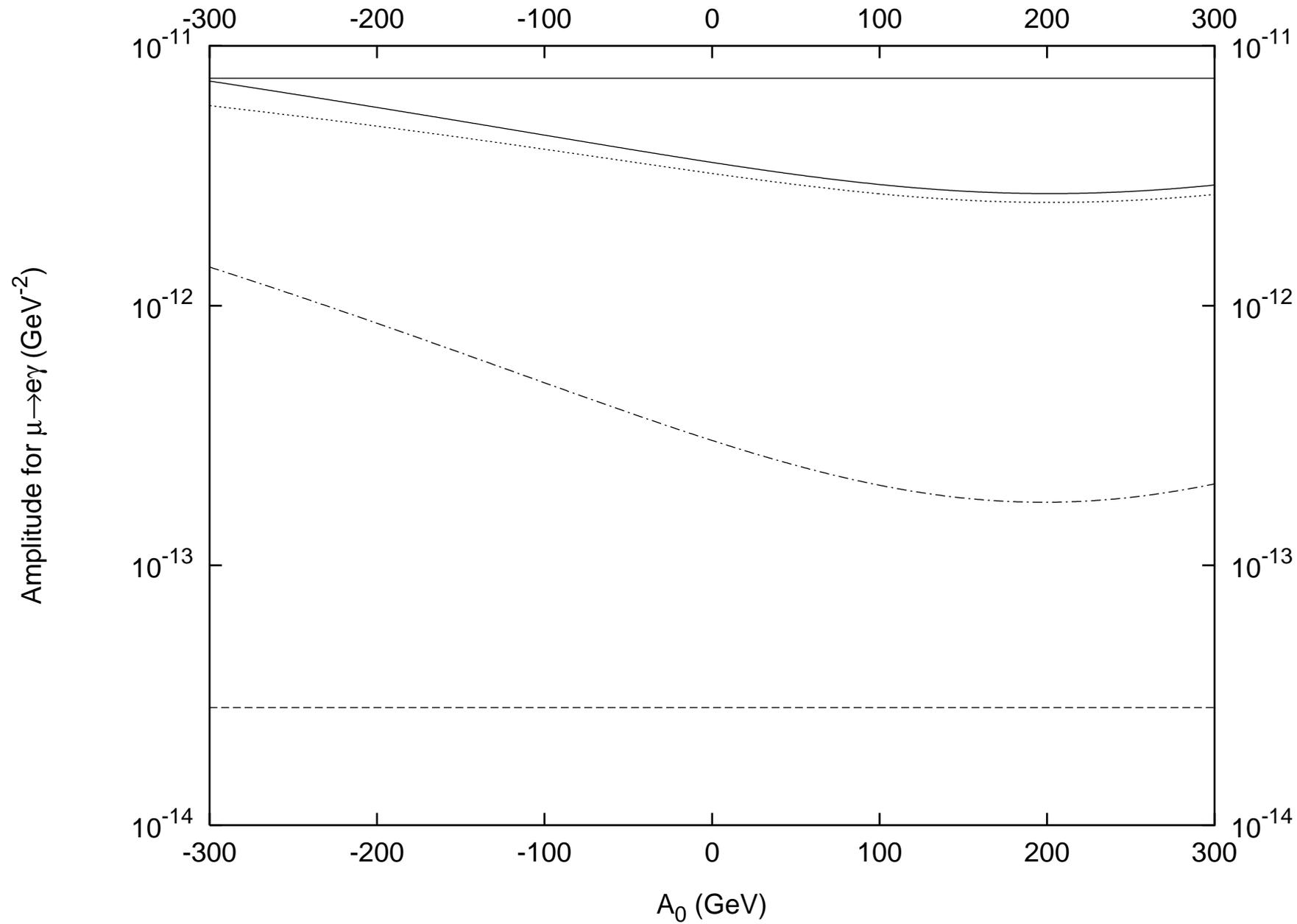
{\bf Figure 11:}
Exactly as for Figure 9, but varying $A_0$ with both $m_0$ and
$M_{1/2}$ set to 100GeV.

\noindent
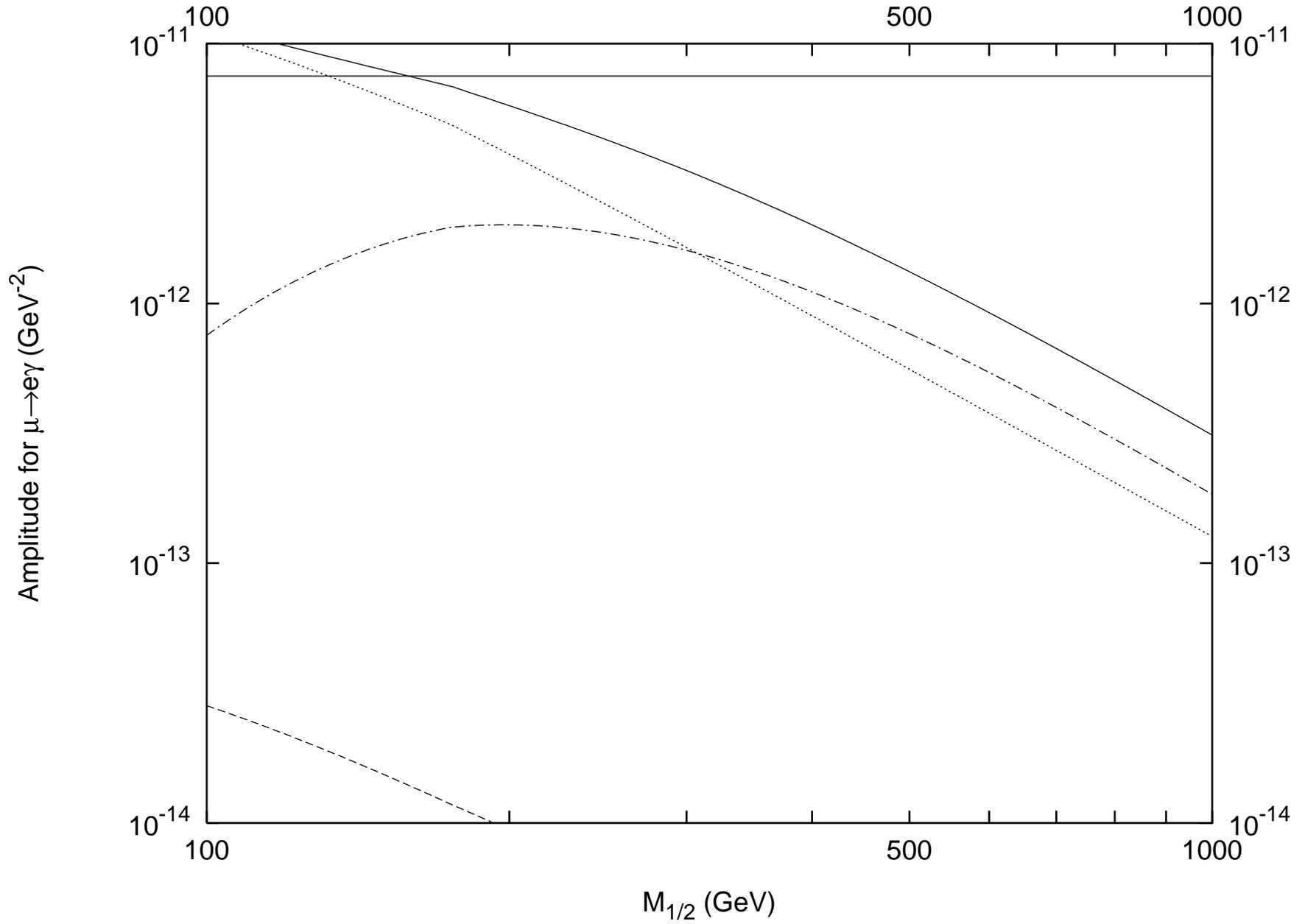
{\bf Figure 12:}
Exactly as for Figure 9, but with $\tan\beta=30$.

\noindent
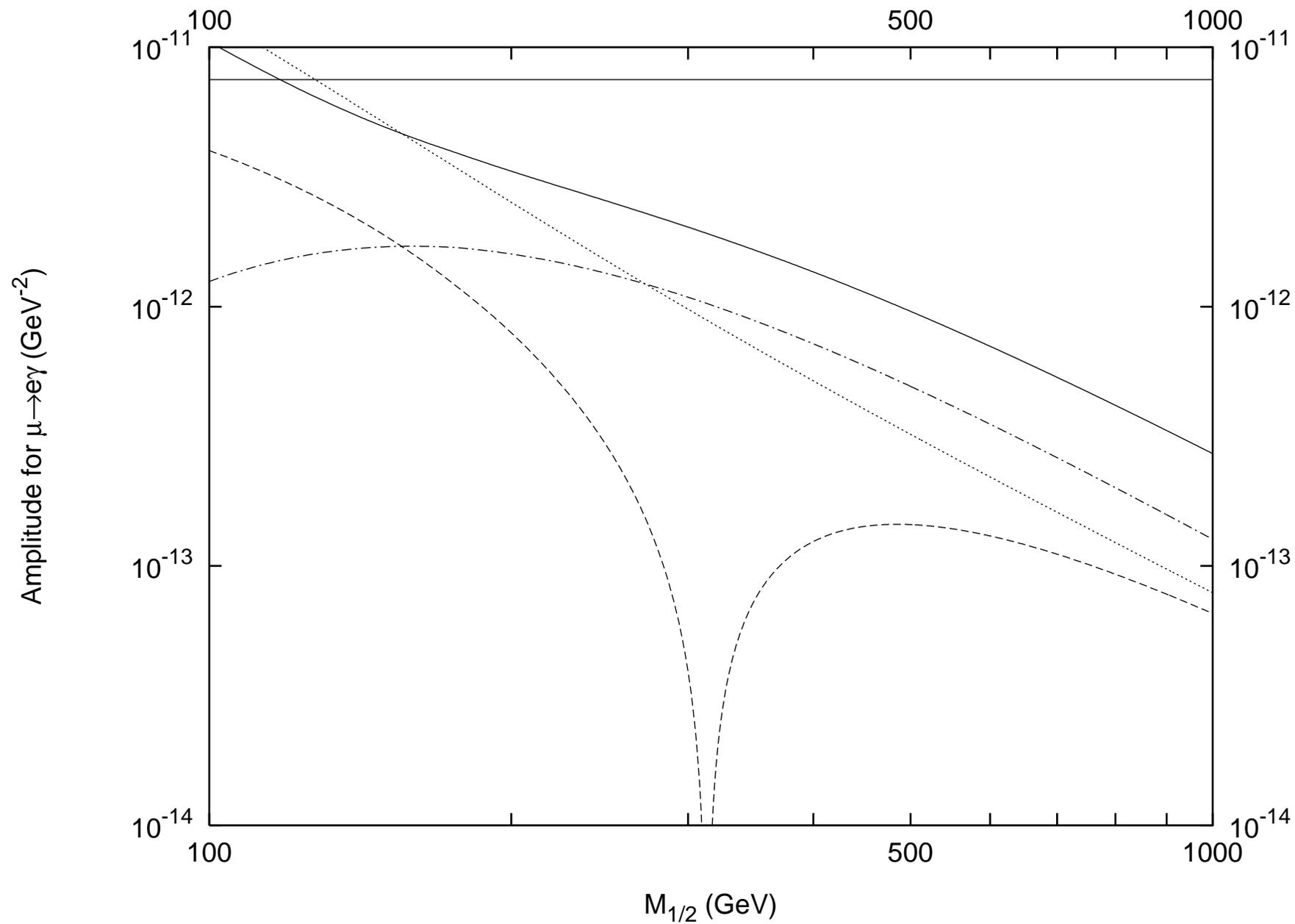
{\bf Figure 13:}
Absolute values of amplitudes for $\mu\to e\gamma$ from direct R-parity
violation diagrams (dashed lines), neutralino mediated diagrams
(dot-dashed lines), and chargino mediated diagrams (dotted lines)
plotted against $M_{1/2}$. We also show the total amplitude (solid
line) and the experimental bound on the amplitude (horizontal solid
line). Parameters are $m_t=175$GeV, $\alpha_3(M_Z)=0.12$,
$\tan\beta=10$, $m_0=100$GeV, $A_0=0$, $\mu_4>0$, and
$\l{131}(M_{GUT})=\l{231}(M_{GUT})=0.01$.

\noindent
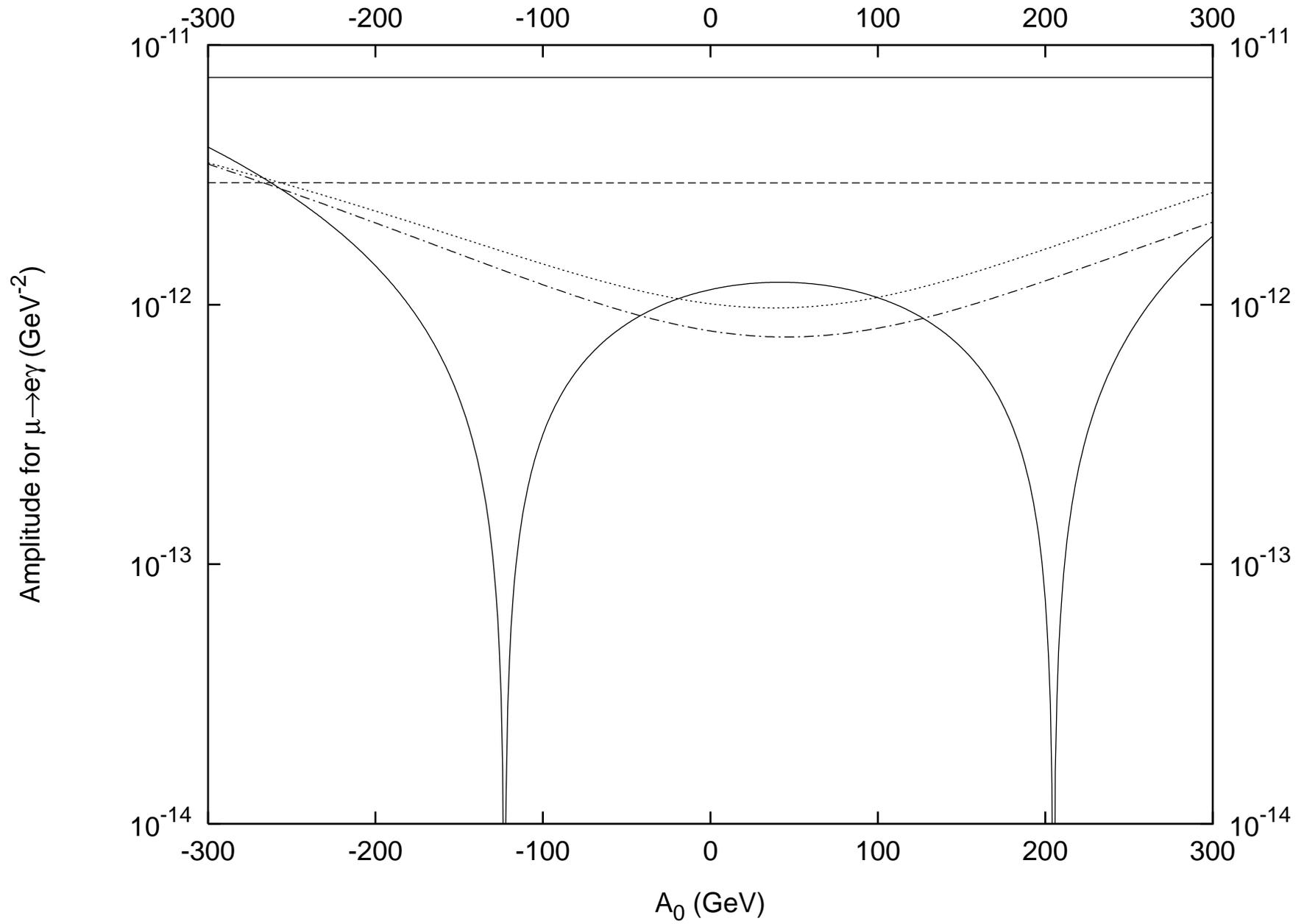
{\bf Figure 14:}
Similar to Figure 13, but varying $A_0$. Parameters are $m_t=175$GeV,
$\alpha_3(M_Z)=0.12$, $\tan\beta=2$, $M_{1/2}=m_0=100$GeV, $A_0=0$,
$\mu_4>0$, and $\l{131}(M_{GUT})=\l{231}(M_{GUT})=0.01$.

\noindent
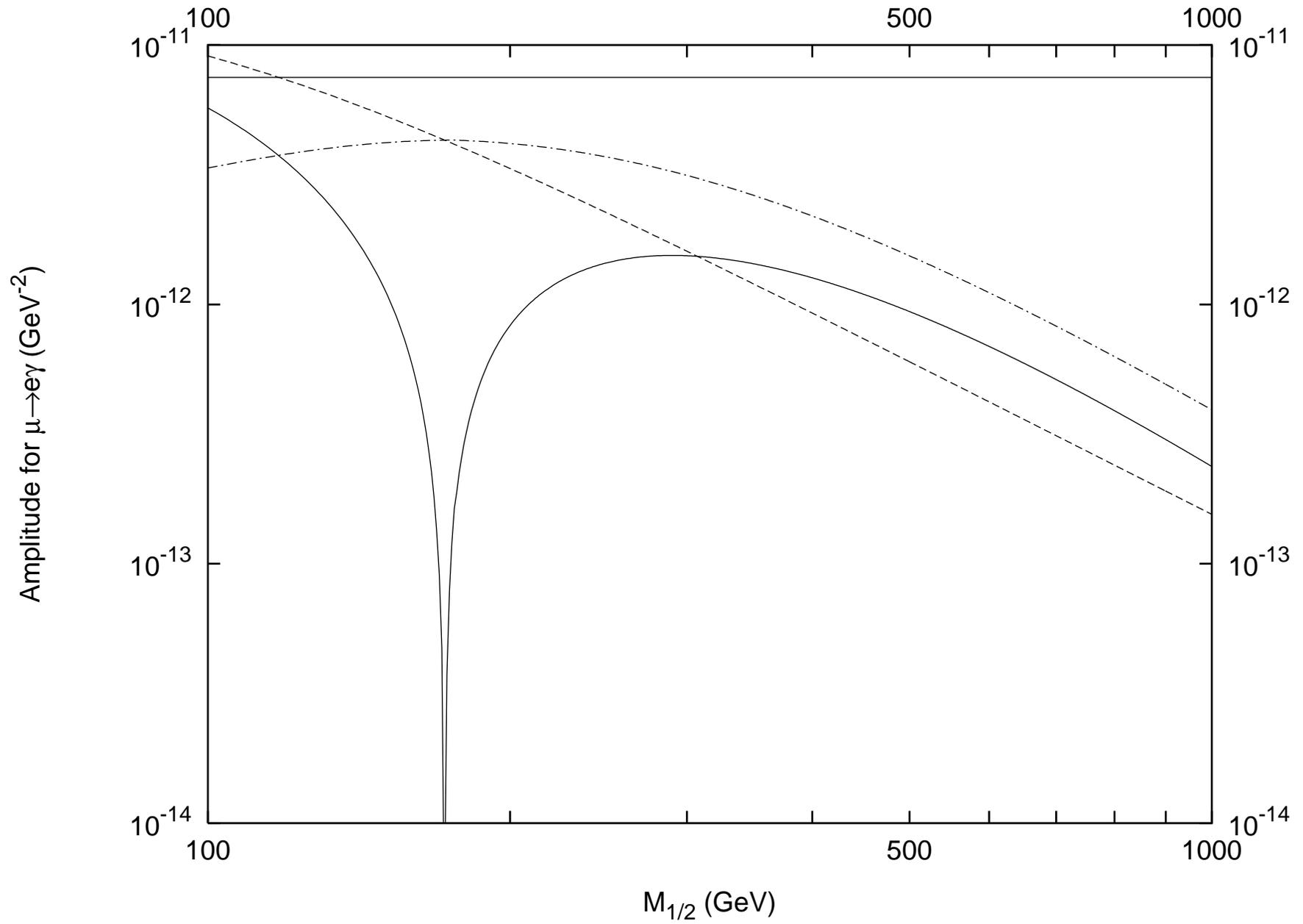
{\bf Figure 15:}
Absolute values of amplitudes for $\mu\to e\gamma$ from direct R-parity
violation diagrams (dashed lines), and neutralino mediated diagrams
(dot-dashed lines) plotted against $M_{1/2}$. We also show the total
amplitude (solid line) and the experimental bound on the amplitude
(horizontal solid line). Parameters are $m_t=175$GeV,
$\alpha_3(M_Z)=0.12$, $\tan\beta=10$, $m_0=100$GeV, $A_0=0$, $\mu_4>0$,
and $\l{121}(M_{GUT})=\l{122}(M_{GUT})=0.01$.


\begin{thebibliography}{99}
\bibitem{revs}
For reviews see for example
H.P.~Nilles, {\it Phys. Rep.} {\bf 110} (1984) 1; \\
H.E.~Haber and G.L.~Kane, {\it Phys. Rep.} {\bf 117} (1985) 75.
%
\bibitem{rpv}
C.S.~Aulah, R.N.~Mohapatra, {\it Phys. Lett.} {\bf B119} (1982) 316; \\
F.~Zwirner, {\it Phys. Lett.} {\bf B132} (1983) 103;\\
S.~Dawson, {\it Nucl. Phys.} {\bf B261} (1985) 297;\\
R.~Barbieri, A. Masiero, {\it Nucl. Phys.} {\bf B267} (1986) 679; \\
S.~Dimopoulos, L.J.~Hall, {\it Phys. Lett.} {\bf B196} (1987) 135.
%
\bibitem{suzuki}
L.J.~Hall, M.~Suzuki, {\it Nucl. Phys.} {\bf B231} (1984) 419.
%
\bibitem{cosmology}
H.~Dreiner, G.G.~Ross, {\it Nucl. Phys.} {\bf B410} (1993) 188.
%
\bibitem{br}
B.~Brahmachari, P.~Roy, {\it Phys.Rev.} {\bf D50} (1994) R39;
and errata in {\it Phys.Rev.} {\bf D51} (1995) 3974.
%
\bibitem{gs}
J.L.~Goity, M.~Sher, {\it Phys. Lett.} {\bf B346} (1995) 69.
%
\bibitem{herbi}
H.~Dreiner, H.~Pois, hep-ph/9511444.
%
\bibitem{roger}
V.~Barger, M.S.~Berger, R.J.N.~Phillips, T.~Wohrmann, hep-ph/9511473.
%
\bibitem{bgh} V.~Barger, G.F.~Giudice, T.~Han,
 {\it Phys. Rev.} {\bf D40} (1989) 2987.  
%
\bibitem{bpw} V.~Barger, R.J.N.~Phillips, K.~Whisnant,
 {\it Phys. Rev.} {\bf D44} (1991) 1629.  
%
\bibitem{bgnn} T.~Banks, Y.~Grossman, E.~Nardi, Y.~Nir,
 {\it Phys. Rev.} {\bf D52} (1995) 5319.
%
\bibitem{analyticRGE}
L.E.~Ib\'a\~nez, C.~L\'opez, {\it Nucl. Phys.} {\bf B233} (1984) 511.
%
\bibitem{ag} K.~Agashe, M.~Graesser, hep-ph/9510439.
%
\bibitem{hagelin}
J.~Hagelin, S.~Kelley, T.~Tanaka, {\it Nucl. Phys.} {\bf B415} (1994) 293; \\
J.~Hagelin, S.~Kelley, T.~Tanaka,
 {\it Mod. Phys. Lett.} {\bf A8} (1993) 2737
%
\bibitem{otherfcnc}
L.J.~Hall, V.A.~Kostelecky, S.~Raby,
 {\it Nucl. Phys.} {\bf B267} (1986) 415; \\
F.~Gabbiani, A.~Masiero, {\it Nucl. Phys.} {\bf B322} (1989) 235; \\
G.C.~Branco, G.C.~Cho, Y.~Kizukuri, N.~Oshimo,
 {\it Phys. Lett.} {\bf B337} (1994) 316;\\
R.~Barbieri, L.~Hall, A.~Strumia,
 {\it Nucl. Phys.} {\bf B445} (1995) 219;
 {\it Nucl. Phys.} {\bf B449} (1995) 437
%
\bibitem{lee}
I.~Lee, {\it Nucl. Phys.} {\bf B246} (1984) 120.
%
\bibitem{rv}
J.C.~Romao, J.W.F.~Valle, {\it Nucl. Phys.} {\bf B381} (1992) 87.
%
\bibitem{vissani}
A.~Smirnov, F.~Vissani, {\it Nucl. Phys.} {\bf B460} (1996) 37.
%
\bibitem{bgmt} R.~Barbieri, M.M.~Guzzo, A.~Masiero, D.~Tommasini,
 {\it Phys. Lett.} {\bf B252} (1990) 251.
%
\bibitem{enqvist}
K.~Enqvist, A.~Masiero, A.~Riotto, {\it Nucl. Phys.} {\bf B373} (1992) 95.
%
\bibitem{hemp}
R.~Hempfling, hep-ph/9511288.
%
\bibitem{pdb}
Particle Data Group Review of Particle Properties, 
 {\it Phys. Rev.} {\bf D50} (1994) 1173.
%
\bibitem{bm} K.S.~Babu, R.N.~Mohapatra,
 {\it Phys. Rev. Lett.} {\bf 75} (1995) 2276.
%
\bibitem{ds} J.-P.~Derendinger, C.A.~Savoy,
 {\it Nucl. Phys.} {\bf B237} (1984) 307.
%
\bibitem{megex}
R.D.~Bolton et al., {\it Phys. Rev.} {\bf D38} (1988) 2077.
%
\bibitem{megsusy}
D.~Choudhury, F.~Eberlein, A.~K\"onig, J.~Louis and
S.~Pokorski, {\it Phys. Lett.} {\bf B342} (1995) 180; \\
S.~Dimopoulos and D.~Sutter, {\it Nucl. Phys.} {\bf B452} (1995) 496;
\\
E.~Gabrielli, A.~Masiero and L.~Silvestrini, hep-ph/9509379.
%
\bibitem{us} B.~de~Carlos, J.A.~Casas and J.M.~Moreno,
 {\it Phys. Rev.} {\bf D53} (1996) 6398.
%
\bibitem{bert} S.~Bertolini, F.~Borzumati, A.~Masiero, G.~Ridolfi,
 {\it Nucl. Phys.} {\bf B353} (1991) 591.
%
\bibitem{gunhab} J.F.~Gunion and H.E.~Haber,
 {\it Nucl. Phys.} {\bf B272} (1986) 1,
 and errata in {\it Nucl. Phys.} {\bf B402} (1993) 567.
%
\bibitem{mv} S.P.~Martin, M.T.~Vaughn,
 {\it Phys. Rev.} {\bf D50} (1994) 2282.
%
\end{thebibliography}
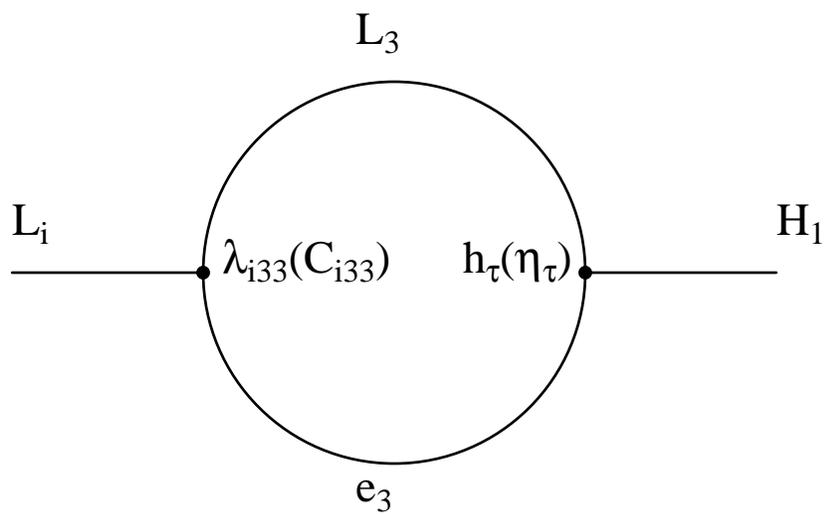
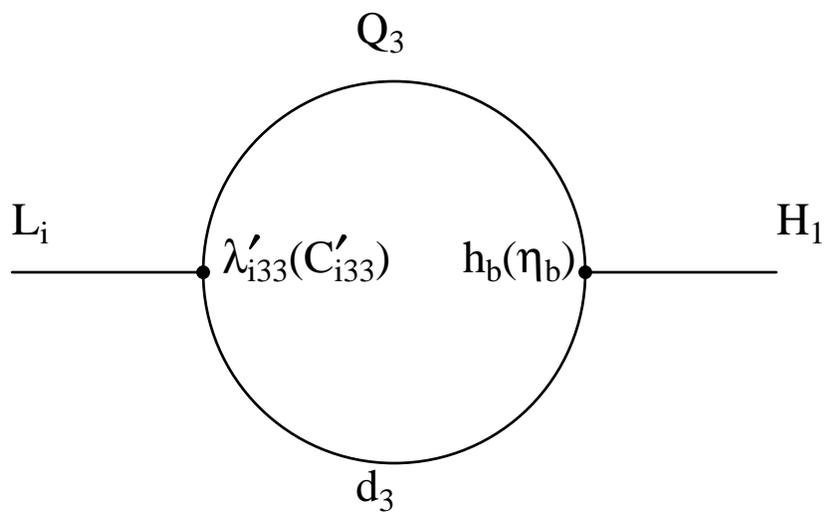
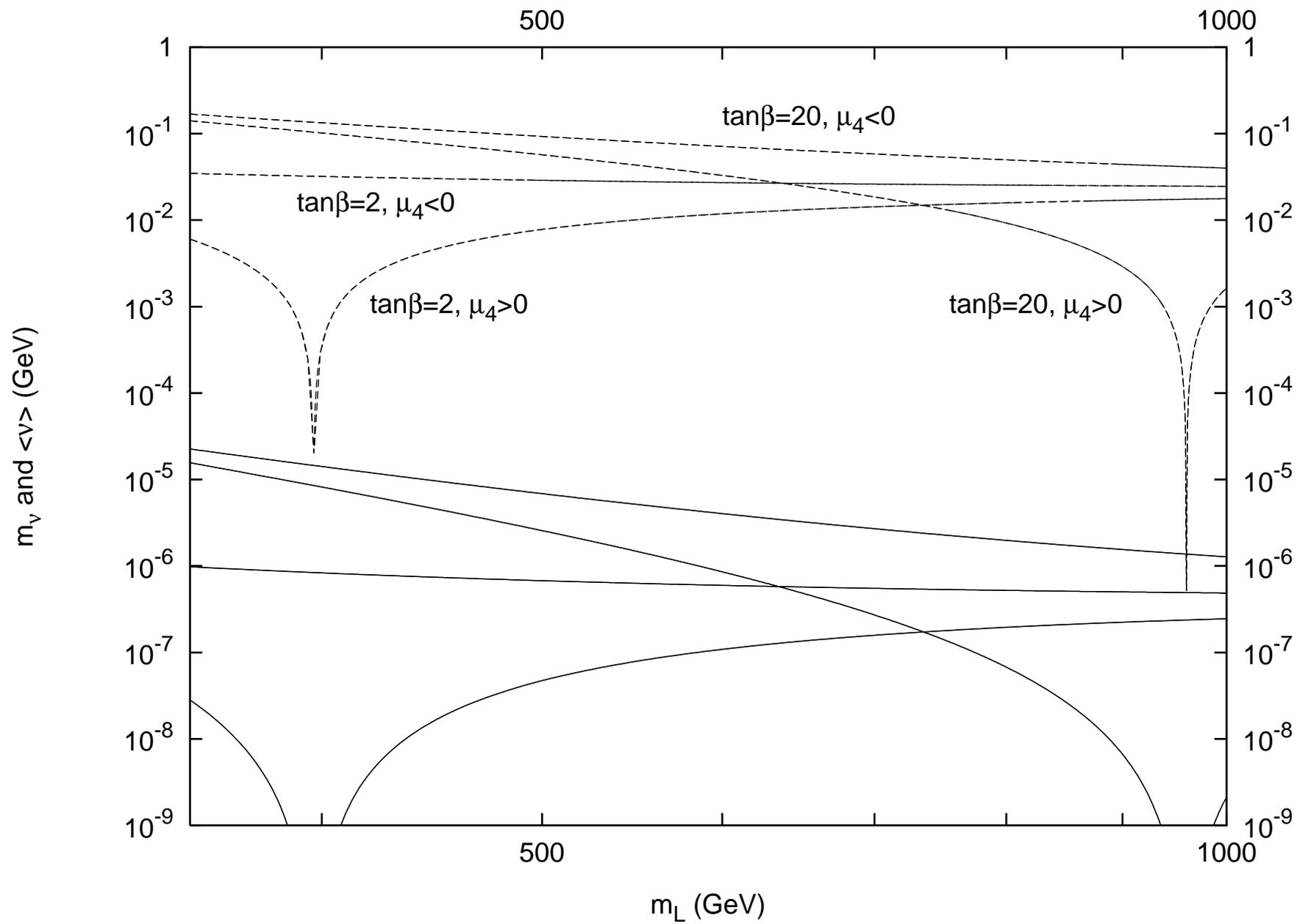
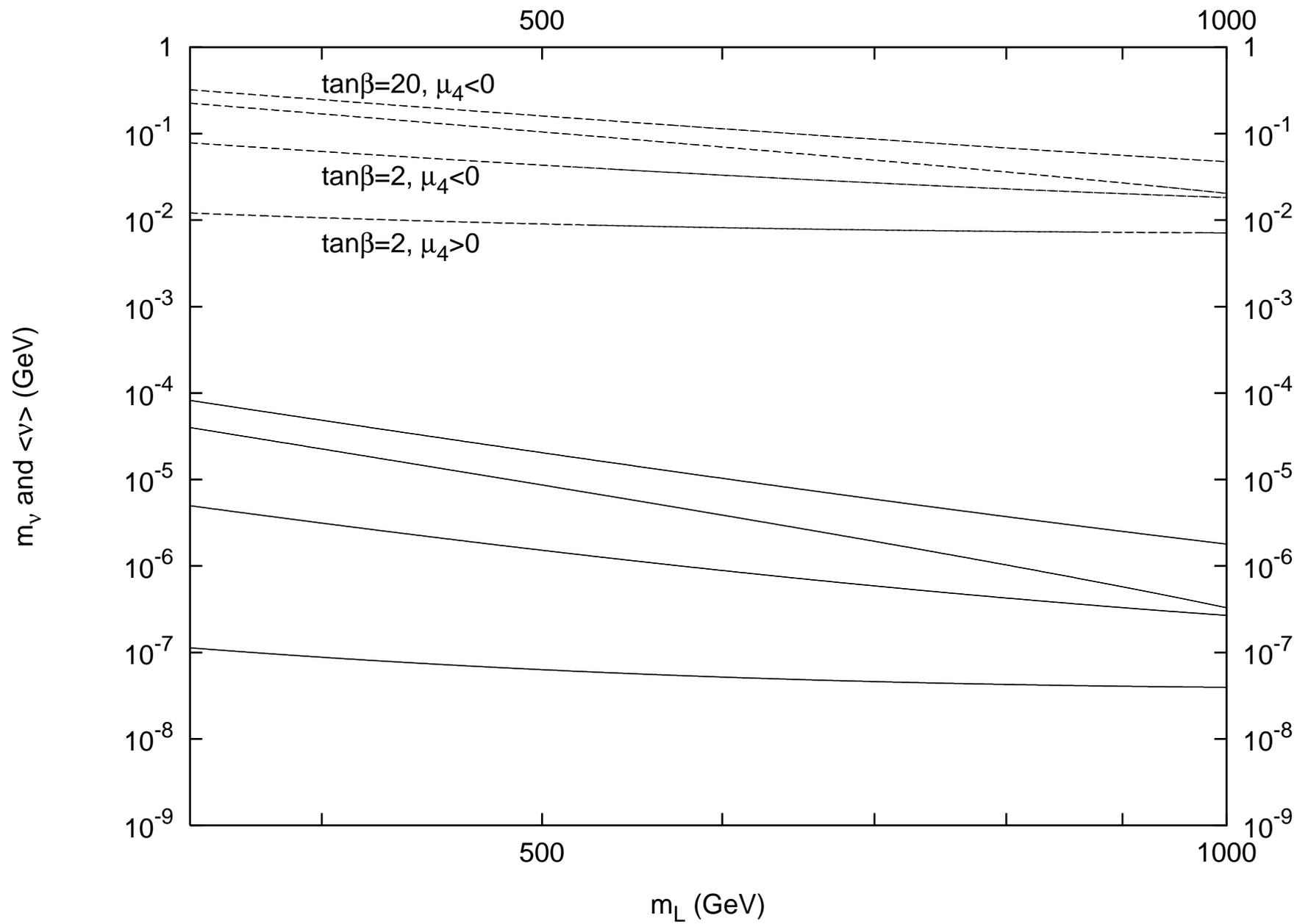
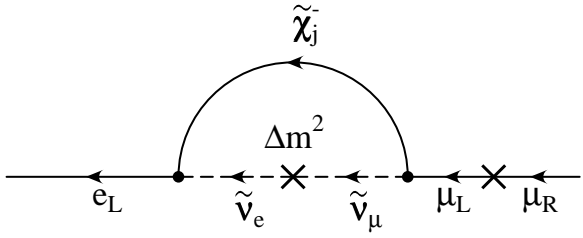
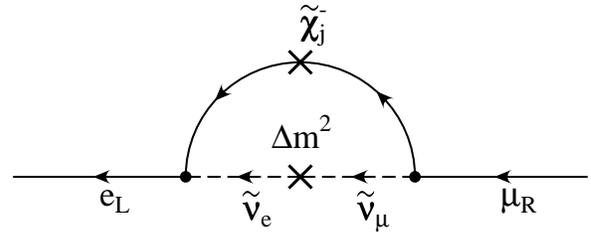
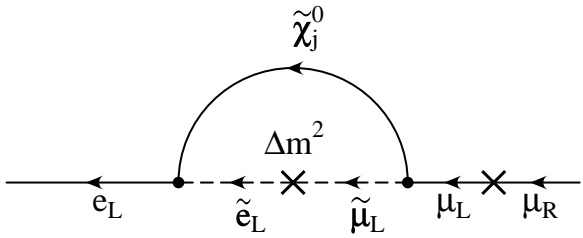
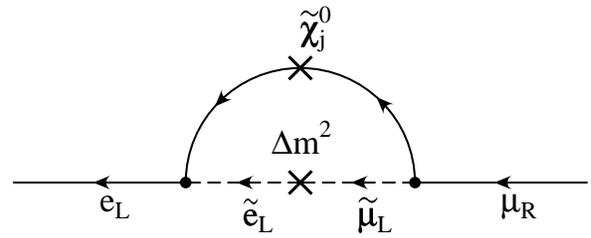
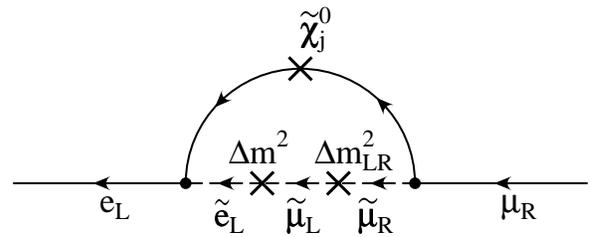
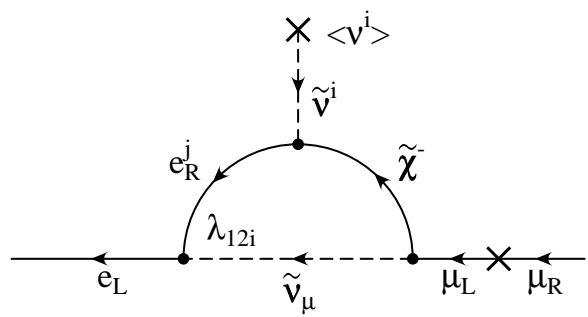
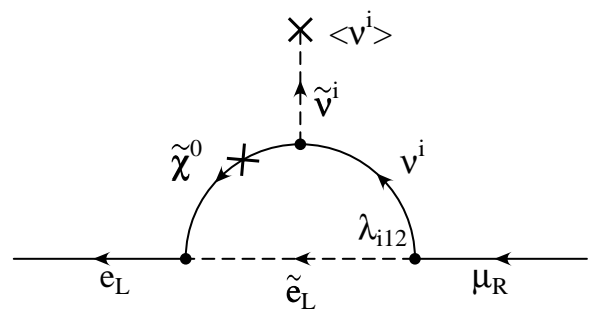
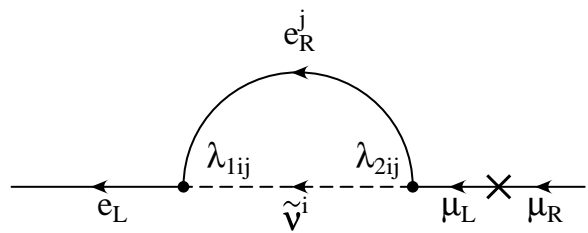
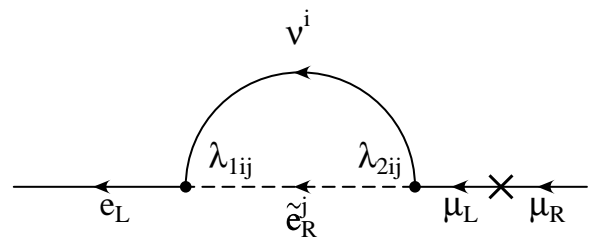
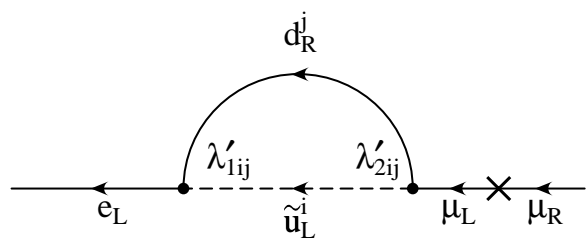
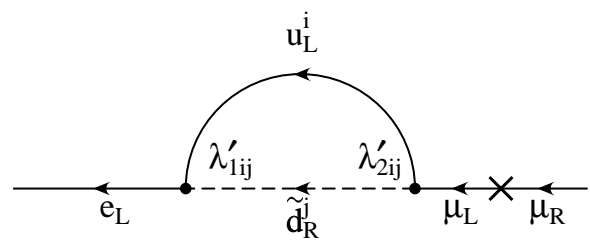
\end{document}